\documentclass[11pt]{article}
\usepackage{jheppub}
\usepackage{amsmath, amsfonts, amssymb, array}
\usepackage{mathrsfs}
\usepackage{slashed}
\usepackage{xcolor}
\newcolumntype{C}[1]{>{\centering\let\newline\\\arraybackslash\hspace{0pt}}m{#1}}
\makeatletter
\g@addto@macro\bfseries{\boldmath}
\makeatother
\usepackage[english]{babel}
\usepackage[autostyle]{csquotes}
\usepackage{float}

\newcommand{\be}{\begin{equation}}
\newcommand{\ee}{\end{equation}}
\newcommand{\bea}{\begin{eqnarray}}
\newcommand{\eea}{\end{eqnarray}}
\newcommand{\nn}{\nonumber}
\usepackage{hyperref}

\newcommand{\ben}[1]{\begin{eqnarray}\label{#1} }
\newcommand{\een}{\end{eqnarray}}

\usepackage{graphicx}
\usepackage{amssymb}
\usepackage{amsthm}
\usepackage{amsmath}
\usepackage{slashed}
\usepackage{color}
\usepackage{bbold}
\usepackage[normalem]{ulem}

\numberwithin{equation}{section}

\makeatletter
\g@addto@macro\bfseries{\boldmath}
\makeatother

\title{Thermodynamics of BPS and Near-BPS AdS$_6$ Black Holes}

\author{Madhu Mishra$^{1}$}
\author{and Amitabh Virmani$^{2}$}

\affiliation{
	$^1$ Indian Institute of Science Education and Research,
	Vithura, Thiruvananthapuram - 695551, India\\
	$^2$ Chennai Mathematical Institute,
H1 SIPCOT IT Park, Kelambakkam, Tamil Nadu 603103, India}

\emailAdd{madhu50315@iisertvm.ac.in}
\emailAdd{avirmani@cmi.ac.in}

\abstract{
 We develop the thermodynamics of BPS and 
 near-BPS AdS$_6$ black holes. We study the phase diagram of BPS black holes in the grand canonical ensemble.
 We highlight two distinct deformations orthogonal to the BPS surface: (i) increasing the temperature  while keeping the charges fixed, (ii) changing the charges while maintaining extremality such that the BPS constraint  is no longer satisfied.  For both these deformations, we show that the considerations of the BPS entropy function 
can be extended 
to describe the near-BPS regime. The excess entropy together with changes in all potentials are perfectly accounted  for via the extremization principle.}

\begin{document}
	\maketitle

\section{Introduction}

The thermodynamics of SU(N) Yang-Mills theory on S$^3 \times \mathrm{R}$ at weak coupling has features similar to AdS physics \cite{Hawking:1982dh, Witten:1998zw, Sundborg:1999ue,  Aharony:2003sx}. However, due to the strong-weak nature of the AdS/CFT duality some qualitative and almost all quantitative aspects are different.   The natural question that arose soon after \cite{Aharony:2003sx} was: can we understand quantitatively the BPS black holes in AdS$_5 \times \mathrm{S}^5$ from 4d ${\cal N}=4$ super-Yang-Mills theory on S$^3 \times \mathrm{R}$? Some important progress was already made more than fifteen years ago \cite{Kinney:2005ej, Romelsberger:2005eg}, but the main question had remained un-resolved.  The growth of indices constructed in \cite{Kinney:2005ej}   was estimated to be $\mathcal{O}(1)$ at large $N$, which does not account  for the black hole entropy. It was interpreted that cancellations between fermionic and bosonic states result in $\mathcal{O}(1)$ growth at large $N$.

 In the last five years, microscopic understanding of BPS black hole entropy in AdS spacetimes has seen impressive progress  \cite{ Benini:2015eyy, Benini:2016rke, Hosseini:2017mds, Hosseini:2018dob, Hosseini:2018usu, Cabo-Bizet:2018ehj,  Choi:2018hmj, Benini:2018ywd, Choi:2019miv, Crichigno:2020ouj};  see \cite{Zaffaroni:2019dhb} for a review (and references therein).  The central insight on the gravity side of the correspondence that led to much of this progress was recasting the entropy together with a certain nonlinear constraint on charges of 5d BPS black holes in terms of the free energy together with a linear  constraint on the (complex) chemical potentials \cite{Hosseini:2017mds}. The complex constraint and the free energy was later derived from the Euclidean quantum gravity considerations \cite{Cabo-Bizet:2018ehj}.  Even a cursory glance at this literature shows that almost all the progress  heavily relies on supersymmetry and techniques special to supersymmetric theories. Thus, on the one hand, it appears that nothing can be said beyond BPS black holes. On the other hand, it is very important in these developments to think of BPS black holes as a limit of a larger family of non-extremal black holes \cite{Cabo-Bizet:2018ehj}. 

In a series of papers, Larsen and collaborators have shown that these recent developments can be ``leveraged'' to describe the near-BPS black holes \cite{Larsen:2019oll, Larsen:2020lhg}.   The proposal is to regard the ``larger space of states'' established in the course of investigating BPS black holes as granted. From this larger space of states, we identify physical states by imposing a constraint for BPS black holes. The key idea of  \cite{Larsen:2019oll, Larsen:2020lhg} is to relax the constraint to accommodate a departure from the BPS limit. Although, a rigorous understanding is lacking,  many properties of near-BPS black holes are captured by this generalisation of the BPS considerations. Larsen and collaborators have studied AdS$_5$ \cite{Larsen:2019oll}, AdS$_4$ and singly rotating AdS$_7$ black holes \cite{Larsen:2020lhg}. In each of the examples studied, the generalisation from BPS to near-BPS works more or less in an identical manner.  This ties up well with the universality discussed in the context of near-AdS$_2$/near-CFT$_1$ correspondence~\cite{Almheiri:2014cka, Kitaev-talks:2015, Maldacena:2016upp, Jensen:2016pah, Engelsoy:2016xyb, Almheiri:2016fws, Larsen:2018iou, Nayak:2018qej, Kolekar:2018sba, Moitra:2018jqs, Iliesiu:2020qvm, Heydeman:2020hhw, Castro:2021fhc, Castro:2021wzn}.  

Charged  AdS$_6$  and AdS$_7$ black holes with multiple rotations are somewhat different from the gravity perspective \cite{Chow:2007ts, Chow:2008ip}. Typically these solutions are most concisely described in Jacobi-Carter coordinates \cite{Chen:2006xh} instead of  the more natural Boyer-Lindquist type coordinates \cite{Gibbons:2004uw}. Due to this extra complication, they remain much less studied compared to their  AdS$_5$ cousins \cite{Chong:2005hr}.
 The aim of this paper is to extend the results  of Larsen and collaborators to the  most general (known) AdS$_6$ black holes. We find it quite remarkable that all considerations of references \cite{Larsen:2019oll, Larsen:2020lhg} find natural generalisation to AdS$_6$ case;  though, there are some non-trivial differences as well.

Our study sheds light on the two parameter reduction to  BPS black holes from the general non-extremal black holes in six-dimensional AdS setting. There are two independent  deformations orthogonal to the BPS surface: i) increasing the temperature   while keeping the charges fixed, ii) changing the charges while maintaining extremality such that a certain BPS constraint is no longer satisfied. For both these deformations, we show that the  previous study of entropy function extremization \cite{Choi:2018fdc} 
 can be extended to describe the near-BPS regime. 
 
 The rest of the paper is organised as follows. 
 
 In section \ref{sec:gravity}, we compute the excess entropy, excess mass, and the changes in all potentials  in going from  the BPS to the near-BPS regime. In section \ref{sec:extremization}, we show that all these quantities are perfectly accounted for via the extremization principle.  We will show that the two independent deformations (i) and (ii)  nicely combine as  a complex parameter. In section \ref{sec:phase_diagram} we study the phase diagram of AdS$_6$ BPS black holes restricting ourselves to the equal rotation case, for simplicity. We close with a brief discussion in section \ref{sec:conclusions}. Other studies of Chow's black holes \cite{Chow:2007ts, Chow:2008ip} include \cite{David:2020ems, Goldstein:2019gpz}.

Results presented in this paper involve much symbolic computations. This would not have been possible without a modern computer algebra system. We have used \verb+Mathematica+ extensively. Files are available on request.

\section{Near-BPS AdS$_6$ black holes}
\label{sec:gravity}
Dimensional reduction of massive type IIA supergravity on S$^4$/$\mathbb{Z}_2$ gives an ${\cal N}=4$ gauged  supergravity in six-dimensions \cite{Cvetic:1999un}. The first charged rotating black hole solution of this gauged supergravity was constructed by Chow in 2008 in \cite{Chow:2008ip}. Chow's general non-extremal solution has mass, two independent angular momenta, and one independent U(1) charge.  These solutions can also be embedded in type IIB theory \cite{Jeong:2013jfc}. In this section, we develop the thermodynamics of these  black holes in the  near-BPS regime. Our presentation closely follows \cite{Larsen:2019oll, Larsen:2020lhg}. 

 \subsection{Charges and potentials}
The physical charges $(M, J_a,  J_b, Q)$  are parametrised by four parameters $(m, a, b, q)$, where $q= 2 m s^2$ with $s = \sinh \delta$.  The physical charges of the general non-extremal black hole are~\cite{Chow:2008ip}:
\begin{align}\label{mass}
M &= \frac{ \pi }{3 G \Xi_a \Xi_b} \left[ 2m \left( \frac{1}{\Xi_a} + \frac{1}{\Xi_b} \right) + q\left(1+\frac{\Xi_a}{\Xi_b} +\frac{\Xi_b}{\Xi_a}\right)\right], \\
\label{charge}
Q &= \frac{2 \pi m s c}{G \Xi_a \Xi_b} , \\ \label{JA}
J_a &= \frac{2\pi m a}{3 G\Xi^2_a \Xi_b}( 1+ s^2\;\Xi_b) , \\ 
\label{JB}
J_b &= \frac{2\pi m b}{3 G\Xi_a \Xi^2_b}( 1+ s^2\;\Xi_a),
\end{align}
where
\begin{align}
 c &= \cosh \delta, & \Xi_a &= 1 - a^2g^2,  &  \Xi_b &= 1 - b^2g^2.
\end{align}
The gauge coupling constant $g$ of the gauged supergravity  is related to the AdS$_6$ length $l_6$ via $g = l_6^{-1}$.  $G$ is the six-dimensional Newton's constant.  We work with the parameter ranges such that  all the conserved charges are non-negative. We take $m \ge 0$, $q\ge 0$ while $ 0 \leq ag < 1$ and $ 0 \leq bg < 1$.

The event horizon of the black hole is located at coordinate $r=r_+$ where  the polynomial equation,
\begin{align}\label{R}
R(r) =(r^2 + a^2) (r^2 + b^2) + g^2[r(r^2 + a^2) + q][r(r^2 + b^2)+q] - 2 m r = 0,  
\end{align}
has its  largest root. After trading $m$ for $r_+$ through  equation \eqref{R}, we can write  thermodynamic 
potentials as functions of the parameters $(r_+,a, b, q)$.  It is often convenient to work with this set of parameters.  The black hole temperature is given as
\begin{align}\label{temp}
T = \frac{2 r_+^2(1+ g^2 r_+^2)(2 r_+^2 + a^2 + b^2) - (1- g^2 r_+^2)(r_+^2 + a^2)(r_+^2 + b^2)+4q g^2 r_+^3 - q^2 g^2}{4 \pi r_+[ (r_+^2 + a^2)(r_+^2 + b^2)+q r_+ ]},
\end{align}
and the electric potential and angular velocities are,
\begin{align}\label{cp}
&\Phi = \frac{2 m s c r_+ }{(r_+^2 +a^2)(r_+^2 + b^2) + q r_+ }, \\
&\Omega_a = \frac{a[ (1 + g^2r_+^2)(r_+^2 + b^2) + q g^2 r_+]}{(r_+^2 +a^2)(r_+^2 + b^2) + q r_+}, \label{OmegaA} \\
&\Omega_b = \frac{b[ (1 + g^2r_+^2)(r_+^2 + a^2) + q g^2 r_+]}{(r_+^2 +a^2)(r_+^2 + b^2) + q r_+}. \label{OmegaB}
\end{align}
The black hole entropy computed from the area law is,
\begin{align}\label{entropy}
S = \frac{2 \pi^2 [ (r_+^2 +a^2)(r_+^2 + b^2) + q r_+]}{3 G \Xi_a \Xi_b}.
\end{align}

\subsection{The BPS limit}
\label{sec:BPS}
The BPS condition for this theory takes the form~\cite{Chow:2008ip}, 
\begin{align}
M_\mathrm{BPS} = \Phi^* Q + \Omega^*_a J_a + \Omega^*_b J_b,
\end{align}
where $\Phi^* = 1$ and $\Omega^*_a  = \Omega^*_b = g$.  All physical configurations must satisfy the BPS bound 
\be
\label{BPS_bound}
M \ge M_\mathrm{BPS}.
\ee  
The general parametric expressions for the physical quantities $(M,Q,J_{a}, J_{b})$ in \eqref{mass}--\eqref{JB} imply,
\begin{align}
& M - Q - g (J_a + J_b)  = - \frac{\pi  g m (a+b)(1-a g) (1-b g)}{6 G e^{2\delta} \; \Xi_a^2\; \Xi_b^2 } \Big( e^{2 \delta} -1 -\frac{2}{(a+b)g}\Big) \times   \nn
\\  & \qquad \qquad  \left( 1 +  a g + b g  + a^2 g^2 - a b g^2 + b^2 g^2 \right) \left( e^{2\delta} + \frac{3 + a g + b g - 
   a^2 g^2 + a b g^2 - b^2 g^2}{1 +  a g + b g  + a^2 g^2 - a b g^2 + b^2 g^2} \right). \nn \\
\end{align}
The expressions in the two parenthesis on the second line are strictly positive in the entire range $0\leq ag < 1, 0\leq bg < 1$,  and $\delta \ge 0$. As a result,  the BPS bound \eqref{BPS_bound} amounts to:
\begin{align}\label{sdelta}
e^{2 \delta} \leq 1 + \frac{2}{(a+b)g}.
\end{align}

 An alternative expression follows from the following  identity,
\begin{align}\label{meq}
(M - g Ja - g Jb)^2 - Q^2 =~& \frac{4 \pi^2}{
 9 \Xi_a^4 \Xi_b^4 G^2} (1 - a g)^2 (1 - b g)^2 ( 2 + (a + b) g)^2 \nn ~\times \\ 
 & \left(m +  \frac{3 + 4 (a+b)g + 3 (a^2 + b^2) g^2  -(a^2 - ab + b^2)^2g^4 } {2 (2+a g + b g)^2} q \right) ~\times \nn \\
 &  \left(m - \frac{1}{2} q g \left(a + b \right) \left( 2 + (a + b) g \right)\right).  \nn\\ 
\end{align}
The expressions in the first two lines on the right hand side are strictly positive in the entire physical range $0\leq ag < 1, 0\leq bg < 1$, $m \ge 0$, $q \ge 0$.  Thus, this identity together with the BPS bound \eqref{BPS_bound} yields the condition on the parameters, 
\begin{align}\label{mbps}
m  \geq \frac{1}{2} q g \left(a + b \right) \left(2 + (a + b) g\right).
\end{align}
Using the definition $q= 2 m s^2$, the equivalence of inequalities \eqref{sdelta} and \eqref{mbps} is easily verified. The inequalities are saturated if and only if the black hole is BPS. 

Imposing the BPS condition, the radial function $R(r)$ takes the form of sum of two squares~\cite{Chow:2008ip}, 
\be \label{R_BPS}
R(r) = \left[ ( 1+ a g + b g) r^2 - a b  \right]^2 + \left[ g r^3 - (a + b + a b g ) r + q g \right]^2.
\ee
The BPS horizon is thus located  at the zero of $R(r)$ in \eqref{R_BPS},
\begin{align}
r^* = \frac{\sqrt{a b}}{\sqrt{1 + a g + b g}}.
\end{align}
Furthermore,
parameter $q$ must take the value, 
\begin{align}
   q^* &= \frac{(a + b) ( 1 + a g) ( 1 + b g)}{g (1+ a g + b g)}\frac{\sqrt{a b}}{\sqrt{1 + a g + b g}}.
\end{align}
We use the superscript $^*$ to denote quantities that take on their BPS values. In the parameter space of general black holes $(r_+,  a, b, q)$, a convenient parameterisation of  BPS black holes is thus, 
\be (r_+ = r^*,  a, b, q=q^*). \ee 

Using the BPS relation  $m  = \frac{1}{2} q g \left(a + b \right) \left(2 + (a + b) g\right)$ from \eqref{mbps}, we also have
\begin{align}
m^* &= \frac{(a + b)^2 ( 1 + a g) ( 1 + b g) (2 + a g + b g)}{2 (1+ a g + b g)}\frac{\sqrt{a b}}{\sqrt{1 + a g + b g}}.
\end{align}

The BPS values of the physical charges take the form \cite{Cassani:2019mms},
\begin{align}
M^* &= \frac{\pi r^* (a + b) (3  +  (a + b) g - (2 a^2 + a b + 2 b^2)g^2    ) }{3 G g (1 - a g)^2 (1 - 
       b g)^2 (1 + (a + b) g)} , \label{BPSM} \\
Q^* &= \frac{\pi r^* (a + b) }{G g (1-a g) (1-b g )}, \label{BPSQ}  \\
J_a^* &= \frac{ \pi  r^*{}^3  (a + b) (1 + a g + 2 b g) }{3 G g b (1-a g)^2 (1-b g) } , \label{BPSJa} \\
J_b^* &= \frac{ \pi  r^*{}^3  (a + b) (1 + 2 a g + b g) }{3 G g  a (1-a g) (1-b g)^2 } \label{BPSJb}, 
\end{align}
and, on the BPS solutions the angular velocities and the chemical potential are, 
\begin{align}
&\Omega_a^* = \Omega_b^* = g, & & \Phi^*=1,& & T^* = 0.&
\end{align}
The  above  quantities saturate the BPS bound \eqref{BPS_bound}: 
$M^* -  g (J_a^* + J_b^*) - Q^* = 0. $ 

The entropy of the BPS black holes,
\begin{align}\label{sentr}
S^* = \frac{2 \pi^2 a b (a + b)}{ 3 G g (1-a g) (1- bg) (1 + a g + b g)},
\end{align}
satisfies the following two relations involving BPS charges \cite{Choi:2018fdc},
\begin{align} \label{BPS_rel_1}
 S^{*3} - \frac{2 \pi^2}{3 g^4 G} S^{*2} - 
 12 \pi^2 \left(\frac{Q^*}{3 g}\right)^2 S^* + \frac{8 \pi^4}{3 g^4 G} J_a^* J_b^* = 0, &  \\
\frac{Q^*}{3 g} S^{*2} + \frac{2 \pi^2}{9 g^4 G} (J_a^* + J_b^*) S^* - 
 \frac{4 \pi^2}{3} \left(\frac{Q^*}{3 g}\right)^3 = 0.&  \label{BPS_rel_2}
\end{align}
From these equations we get a non-linear charge relation -- a BPS charge constraint -- satisfied by the charges of all BPS black holes. Solving for $S^*$  from \eqref{BPS_rel_2} and choosing among the two solutions the manifestly positive solution, we get,
\begin{align} \label{BPS_entropy}
S^* = \frac{\pi}{9 G g^3 Q^*} \left( \sqrt{9 \pi^2 (J_a^* +J_b^*)^2 + 12 g^4 G^2 Q^*{}^4} - 3 \pi (J_a^* +J_b^*) \right).
\end{align}
Inserting this expression into  \eqref{BPS_rel_1} gives the charge relation that all BPS black holes satisfy.

Taking motivation from the resulting expression, we define a ``height'' function $h$,
\begin{align}
\label{height_func}
 h \equiv  - \left( S_h^{3} - \frac{2 \pi^2}{3 g^4 G} S_h^{2} - 12 \pi^2 \left(\frac{Q}{3 g}\right)^2 S_h + \frac{8 \pi^4}{3 g^4 G} J_a J_b\right),
 \end{align}
 where $S_h$ is a auxiliary function of charges, defined as, 
\begin{align} \label{S_h}
 S_h = \frac{\pi}{9 G g^3 Q} \left( \sqrt{9 \pi^2 (J_a +J_b)^2 + 12 g^4 G^2 Q^4} - 3 \pi (J_a + J_b) \right).
 \end{align}
  In the micro-canonical ensemble specified by  angular momenta $J_a^*$, $J_b^*$ and charge $Q^*$, the BPS surface is specified by the vanishing locus of the height function $h$.

We take the BPS entropy a function of physical charges  $S^* (Q^*, J_a^*, J_b^*)$ as written in expression \eqref{BPS_entropy}. This is to contrast with entropy expression \eqref{sentr} written as a function of angular momentum parameters $a, b$. This distinction will play an important role in section \ref{sec:extremization}.

\subsection{Near-BPS black holes}

\label{sec:near_BPS_BH}

We begin with rewriting the function $R(r)$ as a  power series around the position of the BPS horizon $r^*$, 
\begin{align}
R(r) = & - 2 r \left(m -\frac{1}{2} q g (a + b) (2 + (a + b) g ) \right) 
+  A_1 (r-r^*) (q-q^*) 
+ A_2 (r-r^*)^2   \nn \\ &
+ A_3 (r -r^*)^3  
+  A_4 (r-r^*)^4 
+ g^2 (q - q^*)^2 
+ 6 g^2 r^* (r-r^*)^2 (q-q^*)  \nn \\ & 
+ 2 g^2 (r-r^*)^3 (q-q^*)
+  6 g^2 r^* (r-r^*)^5 
+ g^2(r-r^*)^6, \label{R_eq}
\end{align}
where 
\begin{align}
A_1&=-\frac{2}{ab} \left(a+b +
   \left(a^2+b^2\right)g+a b    (a+b) g^2
\right)g  r^{*2} ,\\ 
\label{A_2} 
A_2 &= \frac{r^{*4}}{a^2 b^2}\Big[a^2+6 a b+b^2 + 2 \left(a^3+7 a^2
   b+7 a b^2+b^3\right)g + \left( a^4+14 a^3 b+30
   a^2 b^2+14 a b^3+b^4 \right) g^2  \nn \\  
   &  + 2 a b \left(3
   a^3+7 a^2 b+7 a b^2+3 b^3\right) g^3 + a^2
   b^2 (a+b)^2 g^4 \Big], \\
  A_3 &= 
  \frac{2}{ab} \left( 
2 +3 
   (a+b) g +3
    \left(a^2+4 a b+b^2\right) g^2 + 
   (a+b) \left(2 a^2+a b+2 b^2\right)g^3 \right)r^{*3},\\
  A_4 &= 1 +  \left (a^2 + b^2 +\frac{15 a b}{1 + a g + b g}\right)g^2.
\end{align}

This  looks cumbersome. However, the remarkable fact is that using the above form the horizon equation $R(r_+)=0$  can be organised as the sum of two squares \emph{exactly}:
\begin{align}\label{bps1}
2 r_+  \left(m -\frac{1}{2} q g (a + b) (2+ (a + b) g ) \right) &=B \left[ C^2 +   \frac{g^2}{4\pi^2} D^2 \right],
\end{align}
where
\begin{align}\label{bps2}
C &= T_r (r_+ - r^*) + T_q (q -q^*) + C_2 (r_+ - r^*) ^2 + 
     C_3 (r_+ - r^*) ^3,  \\ 
     \label{bps3}
D &= \frac{2}{q^*} (q-q^*) + D_2 (r_+ - r^*) ^2 + D_3 (r_+ - r^*) ^3.
\end{align}
Notation may appear a bit unwieldy, but it will be clear soon. Comparing the various powers of $(q -q^*)$ and $(r_+ - r^*)$ between \eqref{bps1} and \eqref{R_eq} we get,
\begin{align} \label{B_eq}
B &= \frac{A_2}{T_r^2} =  \frac{4 \pi^2 a^2 b^2 (a+b)^2 (1+ a g)^2 (1+b
   g)^2}{g^2 (1+ g (a+b))^2 A_2 }, &
\end{align}
and
\begin{align}
C_2 &= \frac{A_3}{2 B T_r } , &
D_2 &= \frac{2\pi^2 q^*}{B g^2} \left( 3 g^2 r^* - B C_2 T_q  \right), \\
C_3 &= \frac{1}{2 B T_r}\left(A_4 - B C_2^2 - \frac{g^2}{4 \pi^2}B D_2^2\right), &
D_3 &= \frac{2 \pi^2 q^*}{B g^2} \left(   g^2 - B C_3 T_q  \right).
\end{align}
Expressions for $T_r$ and $T_q$ are given below, \eqref{TR}--\eqref{TQ}.  $B$ introduced in \eqref{B_eq} and $A_2$ introduced in \eqref{A_2}  will feature repeatedly in the expressions below.
 
The quantity $B$ is always non-negative in the range $0 \le ag \le 1, 0 \le b g \le 1$. The left hand side of equation \eqref{bps1} is  non-negative and vanishes exactly when the BPS bound  \eqref{mbps}  is saturated.
Since the right hand side is manifestly a sum of two squares, we see that BPS saturation implies \emph{two conditions} on the black holes: $C= 0,~D =0$.  In the following we analyse these conditions separately.

Let us recall from the previous section that the parametric representation of the BPS limit is $q=q^*$ and $r_+=r^*$. We define near-BPS black holes as those where
\begin{align}
q-q^* \sim r_+-r^* \sim \epsilon.
\end{align}
Identity \eqref{bps1} shows that 
\be
m -\frac{1}{2} q g (a + b) (2+ (a + b) g ) \sim \epsilon^2. 
\ee
This  equation implies  that for near-BPS black holes the parameters $m$ and $q$ are proportional to each other to order $\epsilon$. This observation plays an integral role in the analysis that follows.

Now we establish that for \emph{near-BPS black holes} the conditions $C= 0$ and $D =0$ are precisely the extremality condition $T=0$ and the vanishing of the height function condition $h=0$, respectively.

  \subsubsection{$T \neq 0$ deformation}

BPS black holes have zero temperature: $
T^*= 0.$ The black hole temperature for near-BPS black holes at linear order in the small parameter $\epsilon$ takes the form,
\begin{align}\label{temp2}
T &= T_{r} (r_+ -r^*) + T_{q} (q-q^*) + \mathcal{O}(\epsilon^2).
\end{align}
Our notation is such that  the first term \eqref{bps2} on the right hand side of \eqref{bps1} is proportional to  temperature  \eqref{temp2} at linear order. The derivatives $T_{r}$ and $T_{q}$ in terms of the parameters $a, b, g$ are,
\begin{align}
\label{TR}
 T_r &=  \partial_{r_+} T \Big{|}_{r_+=r^*, \, q=q^*}  =\frac{g (1+ g (a+b)) A_2 }{2 \pi a b (a+b) (1+ a g) (1+b
   g)}, \\  \label{TQ}
 T_{q} &=  \partial_{q} T \Big{|}_{r_+=r^*, \, q=q^*}  = -  \frac{g^2 \left(a+b +  \left(a^2+b^2\right) g+ a
   b  (a+b)g^2\right)}{2 \pi a b (a + 
       b) (1 + a g) (1 + b g) }.
 \end{align}
 Therefore, $C= 0$ condition is the vanishing temperature condition $T=0$ for \emph{near-BPS} black holes.

 $D= 0$ condition for near-BPS black holes is $\delta q := q- q^* = 0$. In this subsection, we take $\delta q =0$ but $T \neq 0$ to first order in $\epsilon$. 
 Since parameters $m$ and $q$ are proportional to each other to order $\epsilon$, the physical charges $Q$, $J_a$, $J_b$  via \eqref{charge}--\eqref{JB} are all proportional to $q$. Since $\delta q =0$, the deformations away from the BPS black holes considered in this subsection do not modify the conserved charges $Q, J_a, J_b$ from their reference values $Q^*, J_a^*, J_b^*$. In particular, the height function constraint $h=0$ continues to be satisfied.

  The response coefficient that characterises the increased temperature is the 
  specific heat 
  \be
  C_T = \frac{dQ}{dT}  = T \frac{dS}{dT},
  \ee
  where the derivative is taken with conserved charges $Q, J_a, J_b$ held fixed. At leading order away from extremality, the specific heat is linear in the temperature so the derivative, \be (\partial_T S)_{Q, J_a, J_b} = \frac{C_T}{T}, \ee is a constant. We next compute this constant.

The entropy $S$ is most conveniently thought of as a function of $(r_+, a, b, q)$. In order to compute $(\partial_T S)_{Q, J_a, J_b}$ we need the matrix of partial derivatives $\frac{\partial(r_+,a, b, q)}{\partial(T, J_a, J_b, Q)}$.
It is straightforward to compute this matrix via,
  \begin{align}
 \left( \frac{\partial(r_+,a, b, q)}{\partial(T, J_a, J_b, Q)}\right)
=  \left( \frac{\partial(T, J_a, J_b, Q)}{\partial(r_+,a, b, q)}\right)^{-1}.  
\end{align}
Having computed these partial derivatives, we can write specific heat as
 \small \begin{align}
 (\partial_T S)_{Q, J_a, J_b}  = \frac{C_T}{T} = &  \left(\frac{\partial{S}}{\partial{r_+}}\right)_{q, a, b} \left(\frac{\partial{r_+}}{\partial{T}}\right)_{Q,J_a, J_b} + \left(\frac{\partial{S}}{\partial{q}}\right)_{r_+,a, b} \left(\frac{\partial{q}}{\partial{T}}\right)_{Q, J_a, J_b} \nn \\ &  +\left(\frac{\partial{S}}{\partial{a}}\right)_{r_+, q, b} \left(\frac{\partial{a}}{\partial{T}}\right)_{Q,J_a,J_b} +\left(\frac{\partial{S}}{\partial{b}}\right)_{r_+,q,a} \left(\frac{\partial{b}}{\partial{T}}\right)_{Q,J_a,J_b}. \label{calc_1}
 \end{align}
We find,
\be \label{ct}
\frac{C_T}{T} = \frac{\pi  \left(1 + 3( a + b) g + (2 a^2 + a b + 2 b^2) g^2\right)r^* B}{3 a b (1+a g)
   (1+ b g)\Xi_a \Xi_b G}.
\ee

This result can be confirmed via the nAttractor mechanism \cite{Larsen:2018iou}. According to the nAttractor mechanism, the elevated temperature is taken into account through the outward displacement of the horizon, via, 
  \begin{align}  \label{calc_2}
  \frac{C_T}{T} = \left(\frac{\partial{S}}{\partial{r_+}}\right)_{q,a,b} \left(\frac{\partial{T}}{\partial{r_+}}\right)^{-1}_{q, {a,b}}, 
  \end{align}
  which again gives \eqref{ct}. Indeed, at the BPS surface,  
\begin{align}
&\left(\frac{\partial{q}}{\partial{T}}\right)_{Q, J_a, J_b} =0, &
&\left(\frac{\partial{a}}{\partial{T}}\right)_{Q,J_a,J_b} =0, &
& \left(\frac{\partial{b}}{\partial{T}}\right)_{Q,J_a,J_b}=0,
   \end{align}
   so the two computations \eqref{calc_1} and \eqref{calc_2} are one and the same.

 As another consistency check, we can arrive at the same expression via a first law consideration. The mass $M$ above the BPS mass $M^*$ can be obtained from equation \eqref{meq}. Inserting the parametric form of the mass excess at the second order from \eqref{bps1}, we obtain 
 \begin{align}
M -M^* = \frac{C_T}{2T} T^2.
\end{align}  
This is perfectly consistent with the first law: 
\be
dM = T dS =  C_T dT = \frac{C_T}{T} (T dT) \implies M -M^* = \frac{C_T}{2T} T^2.
\ee
We also note that $\frac{C_T}{T} $ is strictly positive for $0 \le a g < 1$, $0 \le b g < 1$. 
   
  \subsubsection{$h \neq 0$ deformation}

The near-BPS angular velocities at the linear order are,
\begin{align}\label{exp_om}
& \Omega_a - \Omega_a^* = \Omega_{a {} r}  (r_+ - r^*) 
+  \Omega_{a {} q} (q- q^*) + \mathcal{O}(\epsilon^2), \\
& \Omega_b - \Omega_b^* = \Omega_{b {} r}  (r_+ - r^*) 
+  \Omega_{b {} q} (q- q^*) + \mathcal{O}(\epsilon^2),
\end{align}
and the chemical potential is,
\begin{align}\label{exp_phi}
& \Phi - \Phi^* = \Phi_{r}  (r_+ - r^*)  +  \Phi_{q} (q- q^*) + \mathcal{O}(\epsilon^2),
\end{align}
where the various derivatives at the BPS surface are, 
\begin{align} \label{r-derivatives}
&\Omega_{ar} = \frac{g r^* (1 - a g) \left(  a - b + (a^2 - 4 a b - 3 b^2) g - b (a + b) (a + 2 b) g^2 \right)
}{a b (a + b) (1 + a g) (1 + b g)}, \\
&\Omega_{b r} = \frac{g r^* (1 - b g) \left(  b - a + (b^2 - 4 a b - 3 a^2) g - a (a + b) (b + 2 a) g^2 \right)
}{a b (a + b) (1 + a g) (1 + b g)}, 
\end{align}
\begin{align}  \label{q-derivatives}
&\Omega_{aq} = - \frac{(1-a g)g^2}{(a+b) (1+a g) (1+b g)r^*},  & &\Omega_{bq} = - \frac{(1-b g)g^2}{(a+b) (1+a g) (1+b g)r^*}  \\
 \label{Phi-derivatives}
 &\Phi_{r} =- \frac{g\left( a+b  + (a-b)^2 g -a b  (a+b) g^2 \right) }{r^* (1 + a g) (1 + b g)(1 + 
   (a+b)g)},  & &\Phi_{q} = \frac{g^2}{(1+a g) (1+b g)r^*}  .
\end{align}
Following \cite{Larsen:2019oll, Larsen:2020lhg} we define near-BPS potential $\varphi$ as 
\begin{align}\label{vphi}
\varphi \equiv  3 (\Phi - \Phi^*) -  \frac{1}{g}\Big( (\Omega_a - \Omega_a^*) + (\Omega_b - \Omega_b^*) \Big).
\end{align}
To linear order in the expansion of $\delta r_+ = r_+ - r^* $ and $\delta q =  q -q^*$ we find that 
\begin{align}
\varphi =  \frac{2 }{q^*}(q -q^*). \label{varphi}
\end{align} 
The $r_+ - r^* $ terms cancel out when we consider the combination \eqref{vphi}. From equation \eqref{bps3}, we conclude that at linear order $D = \varphi$.

From equation \eqref{temp2} it follows that the temperature remains zero at linear order when $\delta r_+ = r_+-r^*$ and $\delta q=q-q^*$ are correlated as
\be \label{zero_temp}
 \delta r_+ = - \frac{T_{q}}{T_{r}} \delta  q. 
 \ee
Let us consider deformations where $\delta r_+ $ and $\delta q$ are correlated via \eqref{zero_temp} so the temperature is maintained at zero,  but the black hole becomes non-BPS. 
Inserting the parametric form of the mass excess at the second order from \eqref{bps1} in \eqref{meq} and setting the temperature to zero, we obtain 
 \begin{align}
M -M^* = \frac{C_T}{2T} \frac{g^2}{(2 \pi)^2 } \varphi^2.
\end{align}  

Since the parameters $m$ and $q$ are proportional to each other to order\footnote{As a result, parameter $\delta$ does not change at order  $\epsilon$.} $\epsilon$, the physical charges $Q$, $J_a$, $J_b$  via \eqref{charge}--\eqref{JB} are all proportional to $q$. Thus, shifting $q$ for fixed $a, b$  changes  $Q$, $J_a$, $J_b$ by a common factor. We have around the BPS value,
 \begin{align}
 Q &= Q^* + \delta Q, & J_a &= J_a^* + \delta J_a, & J_b &= J_b^* + \delta J_b , 
\end{align}  
where
 \begin{align}
\delta Q &=  Q^* \frac{\delta q}{q^*}, & \delta J_a &=  J_a^* \frac{\delta q}{q^*}, & \delta J_b &=  J_b^* \frac{\delta q}{q^*}.
\end{align}  
From this variation around the BPS surface $h=0$, we conclude that for near-BPS black holes with $\delta q \neq 0, T = 0$, the height function is proportional to, 
\be
h \propto \delta q \propto \varphi.
\ee
Explicitly, 
\be \label{def_alpha}
 h = \alpha \varphi =  \frac{8 \pi ^6 a^2 b^2 (a+b)^2 (1+a g)
   (1+ b g) \left(1+3 (a+b)g+  \left(2 a^2+a b+2
   b^2\right)g^2 \right)}{27 g^6
   G^3 (1-a g)^3 (1-b g)^3 (1+ (a+b)g )^3
   (1+ (a+b)g + 3 a b g^2)} \varphi.
   \ee
 In the range $0 \le a g <1, 0 \le bg < 1$, the proportionality factor $\alpha$ is non-negative.  Thus, potential $\varphi$ parametrises the possible violation of the constraint on the charges.

\subsection{Near-BPS thermodynamics}
In the previous subsection, we explored the two independent deformations of the BPS configurations. We can now put them together.   Let us begin by considering the mass $M$ above the BPS mass $M^*$. Inserting in  equation \eqref{meq} the parametric mass excess at the second order from equation \eqref{bps1}, temperature from equation \eqref{temp2}, and potential $\varphi$ from equation \eqref{varphi}, we obtain 
 \begin{align}\label{mass_excess}
M -M^* = \frac{C_T}{2T} \left[T^2 + g^2 \frac{\varphi^2}{(2 \pi)^2} \right].
\end{align}  
 This is simply a sum of two independent terms.  Similar formulae for mass excess are also studied in other contexts~\cite{Almheiri:2016fws, Nayak:2018qej, Kolekar:2018sba, Moitra:2018jqs, Iliesiu:2020qvm, Heydeman:2020hhw, Castro:2021fhc}. Note that there is no $T \varphi$ cross-term in the mass excess formula \eqref{mass_excess}. The relative coefficient between the $T^2$ and $\varphi^2$ terms in \eqref{mass_excess} can be argued from AdS$_2$ supersymmetry \cite{Larsen:2019oll}. For our purposes, it is important to note that $\delta q$ contributes to the  temperature for near-BPS black holes.

We next write the near-BPS potentials  in terms of $T$ and $\varphi$. We have, 
\begin{align} \label{near_BPS_potentials1}
& \Omega_a - \Omega_a^* = ( \partial_T \Omega_{a} )  T 
+  (\partial_\varphi \Omega_{a} ) \varphi + \mathcal{O}(\epsilon^2), \\
& \Omega_b - \Omega_b^* = (\partial_T \Omega_{b})  T 
+  (\partial_\varphi \Omega_{b}) \varphi + \mathcal{O}(\epsilon^2), 
\end{align}
and,
\begin{align} \label{near_BPS_potentials2}
& \Phi - \Phi^* = (\partial_T \Phi  ) T 
+ ( \partial_\varphi \Phi  ) \varphi + \mathcal{O}(\epsilon^2). 
\end{align}
The $T$ derivatives are
\begin{align} \label{T_deri}
&\partial_T \Omega_{a} = \frac{\Omega_{ar}}{T_r},&
&\partial_T \Omega_{b} = \frac{\Omega_{br}}{T_r},&
&\partial_T \Phi = \frac{\Phi_r}{T_r},
\end{align}
and the $\varphi$ derivatives are,
\begin{align} \label{phi_deri}
&\partial_ \varphi \Omega_{a} = \frac{q^*}{2} \left( \Omega_{aq} - \frac{T_q}{T_r} \Omega_{ar} \right),& 
&\partial_ \varphi \Omega_{b} = \frac{q^*}{2} \left( \Omega_{bq} - \frac{T_q}{T_r} \Omega_{br} \right), & 
&\partial_ \varphi \Phi= \frac{q^*}{2} \left( \Phi_{q} - \frac{T_q}{T_r} \Phi_{r} \right). 
\end{align}
 Substituting various $q$ and $r$ derivatives from equations \eqref{r-derivatives}--\eqref{Phi-derivatives} we get our final expressions for  $ \Omega_{a,b} - \Omega_{a,b}^*$. We choose not to present those expressions here. We write them in later sections.

 Let us next consider the increase in entropy $S-S^*$ upon changing $r_+$ from $r^*$ to $r^* + \delta r_+$ and $q$ from $q^*$ to $q^* + \delta q$. These perturbations are equivalent to changing $T$ and $\varphi$. Therefore, the change in the entropy $S- S^*$ can be expanded as a sum of terms linear in $T$ and linear in $\varphi$. We find,
\be
S- S^* = \frac{C_T}{T} T + \frac{C_E}{T} \frac{\varphi}{2 \pi}, \label{S_variation}
\ee
where $\frac{C_T}{T}$ was introduced in equation \eqref{ct} and $\frac{C_E}{T}$ is a new response coefficient. It takes the value,
 \begin{align}
\frac{C_E}{T}\Bigg{|}_{S^* = S^*(a,b)} =  \frac{\pi g \left(  a^2 + 4 a b +  b^2 + (a^3 + 2 a^2 b + 2 a b^2 + b^3) g\right)B}{3 a b (a+b) (1+a g + b g)
\Xi_a \Xi_b G}.
\end{align}  

 In the above computation $S$ is taken to be given by expression \eqref{entropy} and $S^*$ is computed at the BPS surface $r=r^*$ and $q=q^*$. For $S^*$ we used expression \eqref{sentr}. From expression \eqref{sentr} we see that upon changing $r_+$ from $r^*$ to $r^* + \delta r_+$ and $q$ from $q^*$ to $q^* + \delta q$,  $S^*$ does not change.

 However, there is an important subtlety here \cite{Larsen:2019oll, Larsen:2020lhg}, which must be properly taken into account.  Indeed, when $S^*$ is thought of as a function of parameters $a, b$, it does not change under $\delta r_+$ and $\delta q$ variations. For thermodynamic considerations,  it is natural to take the BPS entropy to be a function of $Q^*, J_a^*, J_b^*$. While doing so one can consider multiple functions\footnote{Recall that the BPS charges satisfy a non-linear constraint.} of $Q^*, J_a^*, J_b^*$ that give the same two parameter answer \eqref{sentr} for $S^*$.

 As mentioned at the end of section \ref{sec:BPS}, for our purposes we take the BPS entropy to be a function of $Q^*, J_a^*, J_b^*$ via equation \eqref{BPS_entropy}. We  define $C_E$ via equation \eqref{S_variation}.  
Under $\delta q$ variation, 
\begin{align}
 Q^* &\to Q^* + \delta Q, & J_a^* &\to J_a^* + \delta J_a, & J_b^* &\to J_b^* + \delta J_b , 
\end{align}  
where
 \begin{align}
\delta Q &=  Q^* \frac{\delta q}{q^*}, & \delta J_a &=  J_a^* \frac{\delta q}{q^*}, & \delta J_b &=  J_b^* \frac{\delta q}{q^*}.
\end{align}  
This gives an extra contribution, due to the change in $S^*$,
 \begin{align}
\Delta \left[ \frac{C_E}{T} \right] = - \frac{4 \pi^2 g G Q^{*3}}{3 \sqrt{9 \pi^2 (J_a^* +J_b^*)^2 + 12 g^4 G^2 Q^*{}^4} } 
=- \frac{4 \pi ^3 a b (a+b)}{3 g G (1-a g)
   (1-b g) (1 + (a + b) g + 3 a b g^2)},
\end{align}  
such that, 
 \begin{align} \label{CE_final}
\frac{C_E}{T} &= \frac{C_E}{T}\Bigg{|}_{S^* = S^*(a,b)} + \Delta \left[ \frac{C_E}{T} \right]  \\
&= -\frac{\pi g 
\left[ 1 + 3 (a + b) g + (2 a^2 + a b + 2 b^2) g^2 \right] 
\left[ 2 + 3 (a + b) g + (a^2 + 4 a b + b^2) g^2 - a b (a + b) g^3) \right]
B}{3  (a+b) (1+ ag) (1+bg) (1+(a  + b) g) (1 + (a + b) g+ 3 a b g^2)
\Xi_a \Xi_b G}. \nn \label{CE_final2} \\
\end{align}  
In the next section, we will see that this final expression for $\frac{C_E}{T} $ \eqref{CE_final2} is reproduced from the near-BPS extremization principle considerations. 

We note that $\frac{C_E}{T} $ is strictly negative for $0 \le a g < 1$, $0 \le b g < 1$.

We observe from equation \eqref{S_variation} that entropy is proportional to $\varphi$ with negative coefficient 
$\frac{C_E}{T}$ 
for $0 \le a g < 1$, $0 \le b g < 1$.   The thermodynamic stability requires us to focus on the range where the entropy increases. The entropy increases for $\varphi \le 0$. 
This strongly suggests that the physical configuration space is restricted to $\varphi \le 0$. We have not done a careful analysis of the reasons or the implications of this condition. An analysis of the Gibbs free energy for the near-BPS black holes can shed some light on this; we leave such a study for the future. We note that such a study is more involved than the 5d counterpart \cite{Larsen:2019oll, Ezroura:2021vrt}, due to the fact that  relation \eqref{cp} between $\Phi$ and $q$ is non-linear. This is one of the non-trivial features of the six-dimensional non-extremal black holes.

 \section{Near-BPS extremization principle}
 \label{sec:extremization}

Following \cite{Larsen:2019oll, Larsen:2020lhg}, in this section we propose a near-BPS extremization principle for the most general AdS$_6$ black holes. In \cite{Larsen:2019oll, Larsen:2020lhg} similar results are obtained for AdS$_4$, AdS$_5$, and AdS$_7$ black holes. We compare the results from the previous section with the extremization principle considerations. We show perfect agreement.

\subsection{The BPS entropy function} 
We saw in the previous section that the BPS limit of the black hole thermodynamics is a two parameter reduction: $r_+ = r^*$ and $q=q^*$. Since there are two parameters involved, there are multiple ways of reaching the BPS black hole.   In \cite{Cabo-Bizet:2018ehj}\footnote{For earlier work see \cite{Silva:2006xv}.} it was suggested to impose supersymmetry first, followed by extremality.\footnote{Following \cite{Cabo-Bizet:2018ehj} we make a distinction between supersymmetry and extremality. We use the term ``BPS'' to denote a quantity after both supersymmetry and extremality are imposed.} An extremization principle for the entropy of  AdS$_6$ BPS black holes was proposed in \cite{Choi:2018fdc}, where the extremization is done over  supersymmetric configurations. The solution to those extremization equations
is complex; the real part gives the BPS answers. In this subsection we start with a review of the construction of \cite{Choi:2018fdc}; see also \cite{Cassani:2019mms}. This will set the stage for the near-BPS discussion in the following subsections. 

The supersymmetry condition  \eqref{mbps} requires,
\be
q = \frac{2 m}{g (a+b) (2+a g+b g)}. \label{bps_condition}
\ee
The charges of the supersymmetric solutions satisfy, 
\begin{align}\label{sr}
M - \Omega_a J_a - \Omega_b J_b - \Phi Q = 0.
\end{align}
Having restricted to supersymmetric solutions after imposing the supersymmetry condition \eqref{bps_condition}, we can replace $m$  in favor of $r_+$ via the solution of,
\be
R(r_+) = 0,
\ee
where $R(r)$ is given in equation \eqref{R}. We find,
\begin{align} \label{m_r_plus}
m = \frac{1}{2} (a + b) (2 +  (a + b) g) \left[
     (a + b + a b g ) r_+ - g r_+^3 \pm \left( 1 + (a + b)g \right) \sqrt{-(r_+^2 - r^{*2} )^2}\right].
\end{align}

From equation \eqref{m_r_plus}, we note that if we want $m$ to be real, we must take $r_+ = r^*$; that is, we are forced to take the extremal limit in addition to the supersymmetric limit. 
The key idea of \cite{Hosseini:2017mds, Cabo-Bizet:2018ehj} is to impose supersymmetry while staying away from extremality. In that case $m$ must be complex and is given as,
\begin{align}\label{mm}
m = \frac{1}{2} (a + b) (2 +g (a + b) ) (a \mp i r_+) (b \mp i r_+) ( g r_+ \pm i).
\end{align}
We fix the sign to be the  lower signs in this expression,
\begin{align}\label{mm2}
m = \frac{1}{2} (a + b) (2 +g (a + b) ) (a + i r_+) (b + i r_+) ( g r_+ - i).
\end{align}
 The other sign corresponds to sending $i \to - i$. This change straightforwardly propagates to expressions below. 
Accordingly, $q$ must also be complex,
\begin{align}\label{mq}
q= \frac{1}{g} (a + i r_+) (b + i r_+) (g r_+ - i). 
\end{align}

Using expressions for $m$ and $q$ given in \eqref{mm2} and \eqref{mq}, the chemical potentials \eqref{cp}--\eqref{OmegaB} and entropy \eqref{entropy} become 
\begin{align}
& \Phi = \frac{(g r_+ - i) (1+ (a+b)g)}{a b g- i (1+ (a+b)g) r_+} r_+, \\
& \Omega_a = a g \frac{ (g r_+ - i)}{( r_+ - i a) } \frac{(1+ (a + b)g)r_+ + i b}{ (1 +  (a+b)g)r_+ + i a b g}, \\
& \Omega_b = b g \frac{ (g r_+ - i)}{( r_+ - i b) } \frac{(1+ (a + b)g)r_+ + i a}{ (1 +  (a+b)g)r_+ + i a b g}, \\
&S = \frac{2 \pi^2}{3 g \Xi_a \Xi_b } (a + i r_+) (r_+ -i b) \left((1 + a g+b g )r_+ + i a b g \right).
\end{align}
The inverse temperature takes the form,
\begin{align}
& \beta = T^{-1} =   \frac{2 \pi (r_+ - ia ) (r_+ - i b)\left((1 + a g+b g )r_+ + i a b g \right)}{
g (1 + a g + b g) (3 g r_+^2 - 2 i (1 + a g + b g) r_+  -a - b - a b g) (r_+^2-r^{*2})}.
\end{align}
It is checked that these quantities satisfy the constraint,
\begin{align}
\beta \left( 3 g \Phi -  \Omega_a - \Omega_b - g \right) = 2  \pi i.
\end{align}
We note that this condition will not be satisfied if all the chemical potential and the angular velocities were real.

We define new chemical potentials as 
\begin{align}
\label{rcp}
\omega_a &= \beta (\Omega_a - \Omega_a^*),  &\omega_b &= \beta (\Omega_b - \Omega_b^*),   & \Delta & =  \beta (\Phi - \Phi^*).
\end{align}
They take the values
\begin{align} 
\label{omegaA}
\omega_a  &= \frac{2 \pi i  (1-a g) (b+i r_+)}{(3 g r_+^{2} - 2 i (1 + a g + b g) r_+  -a - b - a b g)} , \\
\label{omegaB}
\omega_b &= \frac{2 \pi i (1-b g) (a + i r_+)}{(3 g r_+^{2} - 2 i (1 + a g + b g) r_+  -a - b - a b g)} , \\
\label{Delta}
\Delta  &= -  \frac{2  \pi i (a + i r_+)( b + i r_+)}{(3 g r_+^{2} - 2 i (1 + a g + b g) r_+  -a - b - a b g)}.
\end{align}
The redefined chemical potentials \eqref{rcp} satisfy 
\begin{align}\label{cc}
3 g \Delta - \omega_a - \omega_b  = 2 \pi i.
\end{align}

The chemical potentials $(\omega_a , \omega_b, \Delta)$ remain well defined in the extremal limit $ r_+ \to r^*$.  Since the limit is smooth, these still satisfy the constraint \eqref{cc}: $ 3 g \Delta^* - \omega_a^* - \omega_b^*  = 2 \pi i.$ This condition implies that even in the BPS limit these quantities are complex. This is somewhat surprising, as now we have imposed both supersymmetry and extremality. This observation is closely related to the fact that that the solution to the extremization equations is complex. We will interpret the complex potentials $(\omega_a^* , \omega_b^*, \Delta^*)$ in the following subsection.

To introduce the extremization principle, we begin by considering the quantum statistical relation for black holes. It reads \cite{Gibbons:1976ue},
\begin{align}\label{qsr}
I = -S + \beta (M - \Omega_a J_a - \Omega_b J_b - \Phi Q),
\end{align}
where $I$ is the on-shell action.  The on-shell action $I(\beta, \Omega_a, \Omega_b, \Phi)$ gives the free energy, or equivalently the logarithm of the grand-canonical partition function, as a function of the chemical potentials. Using \eqref{rcp}, (\ref{qsr}), and (\ref{sr}), we obtain the supersymmetric quantum statistical relation \cite{Cabo-Bizet:2018ehj}:
\begin{align} \label{qsr_2}
I = - S - \omega_a J_a - \omega_b J_b - \Delta Q. 
\end{align}
Computing the right hand side of \eqref{qsr_2} for supersymmetric (but not extremal) solutions, we find
\begin{align}
I =  \frac{\pi i}{3 g G} \frac{\Delta^3}{\omega_a \omega_b}. \label{action}
\end{align}
The extremal limit is smooth. Hence, to obtain $I$ for the BPS solutions we simply replace all quantities in equation \eqref{action} with starred quantities.

We wish to emphasise that we have not computed \eqref{action} by evaluating the renormalised on-shell action. 
For Kerr-AdS black holes in arbitrary dimension, a direct calculation of the on-shell actions appears in the appendix of \cite{Gibbons:2004ai}. The analysis is extended for 5d minimal gauged supergravity in \cite{Chen:2005zj}. This result was used in \cite{Cabo-Bizet:2018ehj} to provide a derivation of the extremization principle proposed in \cite{Hosseini:2017mds} from the principles of Euclidean quantum gravity. Along the same lines, a computation of the on-shell action for AdS$_6$ black holes can be used to derive the extremization principle proposed in \cite{Choi:2018fdc}. The supergravity counterterms needed to perform the holographic renormalisation were worked out in \cite{Alday:2014bta}. For a class of non-rotating black holes, related computations have been looked at in \cite{Suh:2018qyv}. As in \cite{Choi:2018fdc, Cassani:2019mms}, we assume that 
the quantum statistical relation \eqref{qsr} is satisfied. It then follows that the on-shell action for supersymmetric (but not extremal) solutions takes the form \eqref{action} \cite{Cassani:2019mms}.

The black hole entropy is defined in the microcanonical ensemble where the conserved charges are specified. The Legendre transform from the grand-canonical ensemble is implemented by the following entropy function for AdS$_6$ black holes \cite{Choi:2018fdc},
\begin{align} 
{\cal S} &= -I - \omega_a J_a - \omega_b J_b - \Delta Q + \Lambda \left( 3 g \Delta - \omega_a - \omega_b  - 2 \pi i \right) \\\label{entropy_function_Choi}
 &= - \frac{\pi i}{3 g G} \frac{\Delta^3}{\omega_a \omega_b} - \omega_a J_a - \omega_b J_b - \Delta Q + \Lambda \left( 3 g \Delta - \omega_a - \omega_b  - 2 \pi i \right).
\end{align}
The extremization over  $ \omega_a , \omega_b, \Delta, \Lambda $ gives the BPS answers. Note that $\Lambda$ is a Lagrange multiplier that enforces the constraint \eqref{cc} on the potentials.

The extremization equations for the entropy function are:
\begin{align}\label{ec1}
& (\partial_{\omega_a} {\cal S})^* = \frac{i \pi \Delta^{*3}}{3 g G \omega^{*2}_a \omega^*_b} - J^*_a - \Lambda^* = 0,   &
& (\partial_\Delta {\cal S})^*   =- \frac {i \pi \Delta^{*2}}{g G \omega_a^* \omega_b^*} +3 g \Lambda^* -Q^* = 0,   \\
& (\partial_{\omega_b} {\cal S})^*  = \frac{i \pi \Delta^{*3}}{3 g G \omega_a^* \omega^{*2}_b} - J_b^* - \Lambda^* = 0,  &
& (\partial_\Lambda {\cal S})^* =3 g \Delta^* - \omega_a^* - \omega_b^*  - 2  \pi i= 0. \label{ec2}
\end{align}
Entropy function at its extremum is, 
\begin{align}\label{ef}
{\cal S}^* =S^* = - 2 \pi i \Lambda^*.
\end{align}

A combination of the extremization conditions \eqref{ec1}--\eqref{ec2} shows that Lagrange multiplier $\Lambda$ satisfies the cubic equation, 
\be
(Q^* - 3 g \Lambda^*)^3 + \frac{ 9 i \pi }{g G} (\Lambda^* + J_a^*) (\Lambda^* + J_b^*)   = 0. 
\ee
We can write this equation equivalently as 
\be
\label{ce}
 \Lambda^{*3} + c_1 \Lambda^{*2} + c_2 \Lambda^* + c_3 = 0,
\ee
where the coefficients $c_1, c_2$ and $c_3$ are given as follows:
\begin{align}
c_1 & =  -\frac{Q^*}{g} - \frac{i \pi}{3 g^4 G}  , & 
c_2 & = \frac{g^2 G Q^{*2} - i \pi (J_a^* +J_b^*)}{3 g^4 G}  , & 
c_3 &= - \frac{g G Q^{*3} + 9 i \pi J_a^* J_b^*}{27 g^4 G} . 
\end{align}

 The requirement that the BPS entropy \eqref{ef} be real demands that $\Lambda^*$ must be purely imaginary. Substituting \eqref{ef} in \eqref{ce} gives an equation involving $S^*$. Taking $Q^*, J_a^*, J_b^*, S^*$ to be all real, the  real and imaginary parts of that equation gives two independent equations. The real part gives:
\begin{align}\label{mS}
\frac{Q^*}{3 g} S^{*2} + \frac{2 \pi^2}{9 g^4 G} (J_a^* + J_b^*) S^* - 
 \frac{4 \pi^2}{3} \left(\frac{Q^*}{3 g}\right)^3 = 0,& 
\end{align}
which is precisely the relation  \eqref{BPS_rel_1}. 
The imaginary part gives
\begin{align} 
 S^{*3} - \frac{2 \pi^2}{3 g^4 G} S^{*2} - 
 12 \pi^2 \left(\frac{Q^*}{3 g}\right)^2 S^* + \frac{8 \pi^4}{3 g^4 G} J_a^* J_b^* = 0,
\end{align}
which is precisely the relation  \eqref{BPS_rel_2}. 
The positive solution of  equation  \eqref{mS} gives the BPS entropy \eqref{BPS_entropy}. The value of $\Lambda$ via \eqref{ef} is:
\begin{align} \label{Lambda}
\Lambda^* =  - \frac{S^*}{2 \pi i} = \frac{i}{18 G g^3 Q^*} \left( \sqrt{9 \pi^2 (J_a^* +J_b^*)^2 + 12 g^4 G^2 Q^*{}^4} - 3 \pi (J_a^* +J_b^*) \right).
\end{align}

\subsection{Complex potentials}
\label{sec:complex_potentials}
From the extremization equations we can also compute the BPS potentials $(\omega_a^*, \omega_b^*, \Delta^*)$. A convenient way to do so is as follows. We describe the procedure for computing $\omega_a^*$. For  $\omega_b^*$ and $\Delta^*$ the discussion is very similar. To find $\omega_a^*$ we first eliminate $\Delta^*$ between the two equations $(\partial_{\omega_a} {\cal S})^*=0$ and $(\partial_{\omega_b} {\cal S} )^*=0$. This gives
\be
\omega^*_b =   \omega_a^* \frac{J_a^*+ \Lambda^*}{J_b^*+ \Lambda^*} . \label{first_elimination}
\ee
Next we eliminate $\omega_b^*$ between the two equations $(\partial_{\Delta} {\cal S})^*=0$ and $(\partial_{\omega_a} {\cal S})^*=0$. This gives, 
\be
\Delta^* =  3 \omega_a^* \frac{J_a^* + \Lambda^*}{ 3 g \Lambda^* - Q^*} .  \label{second_elimination}
\ee
Substituting expressions \eqref{first_elimination} and \eqref{second_elimination} in $(\partial_\Lambda S)^*=0$ gives
\begin{align} \label{omega_a_BPS_choi}
\omega_a^* &=\frac{  2 \pi i(J_b^* + \Lambda^* )( 3 g \Lambda^*- Q^*)}
{9 g (J_a^* + \Lambda^*) (J_b^* + \Lambda^*) - ( 3 g \Lambda^*- Q^*) (J_a^* + J_b^* + 2 \Lambda^*) }.
\end{align}
Similarly,
\begin{align} \label{omega_b_BPS_choi}
\omega_b^* &=\frac{  2 \pi i(J_a^* + \Lambda^* )( 3 g \Lambda^*- Q^*)}
{9 g (J_a^* + \Lambda^*) (J_b^* + \Lambda^*) - ( 3 g \Lambda^*- Q^*) (J_a^* + J_b^* + 2 \Lambda^*) },\\ 
\Delta^* &= \frac{ 6 \pi i (J_a^* + \Lambda^* ) (J_b^* + \Lambda^* )}
{9 g (J_a^* + \Lambda^*) (J_b^* + \Lambda^*) - ( 3 g \Lambda^*- Q^*) (J_a^* + J_b^* + 2 \Lambda^*) }.
\label{Delta_BPS_choi}
\end{align}

Upon using the parameterisation \eqref{BPSQ}--\eqref{BPSJb} for $Q^*, J_a^*, J_b^*$ we get $\omega_a^*$,  $\omega_b^*$, and $\Delta^*$ in the form \eqref{omegaA}--\eqref{Delta} with $r_+ $ replaced with $r^*$. The real parts of the these equations give the derivatives of the original potentials $(\Omega_a, \Omega_b, \Phi)$ with respect to the temperature:
\begin{align}
\text{Re} \; \omega_a^* &= - \frac{(1 -a g) \left( b - a + (3 b^2 + 4 a b - a^2) g + b (2 b^2 + 3 a b + a^2) g^2 \right)g^2  B}{2\pi  a b r^* (a + b)^2 (1 + a g)^2 (1 + b g)^2} =\partial_T \Omega_a, \\
\text{Re} \; \omega_b^* &= - \frac{(1 - b g) \left( a - b + (3 a^2 + 4 a b - b^2) g + a (2 a^2 + 3 a b + b^2) g^2 \right)g^2  B}{2\pi  a b r^* (a + b)^2 (1 + a g)^2 (1 + b g)^2} =\partial_T \Omega_b, \\
\text{Re}\; \Delta^* &= - \frac{\left(a + b + (a - b)^2 g - a b (a + b) g^2 \right)g^2  B}{2\pi  a b r^* (a + b) (1 + a g)^2 (1 + b g)^2} = \partial_T \Phi.
\end{align}
The imaginary parts give the derivatives of the original potentials with respect to $\varphi$:
\begin{align}
\text{Im} \; \omega_a^* &= - \frac{g^2 (1 - a g) (1 + a g + b g) (3 a + b + a b g + b^2 g) B}{2 \pi a^2 b (a + b)^2 (1 + a g) (1 + b g)^2 } = \frac{2 \pi}{g} \partial_\varphi \Omega_a , \\ 
\text{Im} \; \omega_b^* &= - \frac{g^2 (1 - b g) (1 + a g + b g) (3 b + a + a b g + a^2 g) B}{2 \pi  a b^2 (a + b)^2 (1 + a g)^2 (1 + b g) } = \frac{2 \pi}{g} \partial_\varphi \Omega_b , \\ 
\text{Im} \; \Delta^* &=  \frac{g^2     \left(2 + 5 (a + b) g + (3 a^2 + 4 a b + 3 b^2) g^2 + 
   a b (a + b) g^3\right)B}{2 \pi a b (a + b)  (1 + a g)^2 (1 + b g)^2} =\frac{2 \pi}{g} \partial_\varphi \Phi.
\end{align}
The matching of these expressions from the black hole side \eqref{T_deri} and \eqref{phi_deri} is remarkable. This is also the case for AdS$_d$, $d=4,5$ \cite{Larsen:2019oll, Larsen:2020lhg}. Our analysis extends the work of Larsen et.~al.~\cite{Larsen:2019oll, Larsen:2020lhg} to  the most general black holes in $d=6$. Some  results in  $d=7$  were reported in \cite{Larsen:2020lhg}. In $d=7$ the most general known black hole solution has three-independent rotation parameters and one charge \cite{Chow:2007ts}. It is desirable to generalise the analysis of \cite{Larsen:2020lhg} to these black holes.\footnote{We thank David Chow and Bidisha Chakrabarty for discussions on this problem. }

Note that the analysis of this subsection gives the complete near-BPS potentials \eqref{near_BPS_potentials1}--\eqref{near_BPS_potentials2}.

\subsection{Near-BPS extremization principle}

\label{sec:near_BPS_extremization_principle}

In this subsection, we introduce a near-BPS extremization principle that accounts for the near-BPS entropy. We need to relax the constraint from its strict BPS version. Following \cite{Larsen:2019oll, Larsen:2020lhg} we consider,
\begin{align}
3 g (\Phi -\Phi^*) - (\Omega_a - \Omega_a^*) -  (\Omega_b - \Omega_b^*) = g  \varphi + 2  \pi i T. \label{new_cc}
\end{align}
For $T = 0$ this condition is just the definition of the near-BPS Potential $\varphi$ introduced in (\ref{varphi}). For $\varphi =0$ this condition is equivalent to the constraint \eqref{cc} in the extremal limit $T \rightarrow 0$. As we saw in the previous section, both $T \neq 0$ and $\varphi \neq 0$ take us away from the BPS surface.  
The prescription of \cite{Larsen:2019oll, Larsen:2020lhg} tells us that once the physical parameter are such that $T=0, \varphi \neq 0$, the non-zero temperature can be taken into account by the substitution $g \varphi \to g \varphi + 2 \pi i T$.

We consider the entropy function $\mathcal{S}$,
\begin{align} \label{entropy_func_non_BPS}
T \mathcal{S} &= -\frac{\pi i}{3 g G} \frac{(\Phi- \Phi^*)^3}{(\Omega_a - \Omega_a^*)(\Omega_b - \Omega_b^*)}  - (\Phi -\Phi^*) Q - (\Omega_a - \Omega_a^*) J_a - (\Omega_b - \Omega_b^*) J_b \nn \\
&+ \Lambda \left( 3 g (\Phi -\Phi^*) - (\Omega_a - \Omega_a^*) -  (\Omega_b - \Omega_b^*) - g \; \varphi - 2 i \pi T \right).
\end{align}
The key change from the supersymmetric case is the relaxation of the constraint from \eqref{cc} to  \eqref{new_cc}. The extremization is now to be done with respect to $\Phi$, $\Omega_a$, $\Omega_b$ and $\Lambda$. The extremum value of the entropy function $\mathcal{S}$ is
\begin{align}
T \mathcal{S} = - \Lambda (  g \varphi + 2 \pi i T). 
\end{align}

Once again, a combination of the extremization equations shows that the Lagrange multiplier $\Lambda$ satisfies the same cubic equation \eqref{ce}. To keep the notation simple, we do not use any special subscript or superscript to denote the quantities $Q, J_a, J_b, \Lambda$ at the extremum value of $\mathcal{S}$. The cubic equation takes the form,
\be 
(Q - 3 g \Lambda)^3 + \frac{ 9 i \pi }{g G} (\Lambda + J_a) (\Lambda + J_b)   = 0. 
\ee
which we can write equivalently as, 
\be
\label{ce2}
 \Lambda^{3} + c_1 \Lambda^{2} + c_2 \Lambda + c_3 = 0,
\ee
where the coefficients $c_1, c_2$ and $c_3$ are,
\begin{align}
c_1 & =  -\frac{Q}{g} - \frac{i \pi}{3 g^4 G}  , & 
c_2 & = \frac{g^2 G Q^{2} - i \pi (J_a +J_b)}{3 g^4 G}  , & 
c_3 &= - \frac{g G Q^{3} + 9 i \pi J_a J_b}{27 g^4 G} . 
\end{align}

The key difference between the BPS and the near-BPS cases is that the near-BPS does not guarantee  a purely imaginary solution of the cubic equation \eqref{ce2}. We can write the cubic equation in the form, 
\begin{align}
  \left(\Lambda - A \right) \left(\Lambda^2 + B \Lambda + C \right) = -  \frac{ i }{8 \pi^3} h
\end{align}
where on the right hand side is the height function \eqref{height_func}, which in general is non-vanishing. The coefficients $A$, $B$, and $C$ are
\begin{align}
A &= \frac{i}{2\pi} S_h, &
B &=  c_1 + A, &
C &= c_2 +  A B,
\end{align}
where $S_h$ is defined in \eqref{S_h}. There are no approximations here; it is simply a rewriting of equation \eqref{ce2} in a convenient form. 

In this new form, it is manifest that when the height function is zero, the root $\Lambda = A$ gives the entropy that we already found in the BPS case \eqref{Lambda}. For small violations of the constraint $h = 0$, we can perturb around the root $\Lambda = A$ and find the shift proportional to $h$. The requisite root is at $ \Lambda = A + \delta \Lambda$ where
\begin{align}
\delta \Lambda = - \frac{i}{8 \pi^3} \frac{ h}{\left[ A^2 + A B  + C \right]}.
\end{align}

In the near-BPS case, the height function is related to $\varphi$ at linear order as in equation \eqref{def_alpha},
\be
h = \alpha \varphi.
\ee
The non-zero temperature is taken into account through $g \varphi \to g \varphi + 2 \pi i T$. As a result, according to the prescription of \cite{Larsen:2019oll, Larsen:2020lhg},
\be
S - S^* = - \text{Re} \;  \left[ 2 \pi i \delta \Lambda \Big{|}_{g \varphi \to g \varphi + 2 \pi i T}\right]. \label{larsen_main_prescription}
\ee
Substituting in values of various parameters $A, B, C, \alpha$ etc, and the parameterisation \eqref{BPSQ}--\eqref{BPSJb} for the BPS charges, we find 
\be
S- S^* = \frac{C_T}{T} T + \frac{C_E}{T} \frac{\varphi}{2 \pi}, \label{S_variation_2}
\ee
with $\frac{C_T}{T}$ given in \eqref{ct} and $\frac{C_E}{T}$ given in \eqref{CE_final2}. 
 The shift in entropy $S- S^*$ computed from black hole thermodynamics, namely equation \eqref{S_variation}, precisely matches with equation \eqref{S_variation_2}.
The match between these rather complicated functions computed using two rather different methods in the near-BPS regime is astonishing.

\subsection{Near-BPS potentials}
We can also compute the near-BPS potentials $(\Omega_a, \Omega_b, \Phi)$ from the near-BPS extremization equations.
 The computation proceed parallel to subsection \ref{sec:complex_potentials}. The real part of each potential becomes a linear combination of $\varphi$ and $T$. For the angular velocities we find,
\begin{align}
\text{Re} \: (\Omega_a - \Omega_a^*) = &~\text{Re} \: \left[\frac{  (J_b^* + \Lambda^* )( 3 g \Lambda^*- Q^*) \left( \varphi g + 2 \pi i T\right)}
{9 g (J_a^* + \Lambda^*) (J_b^* + \Lambda^*) - ( 3 g \Lambda^*- Q^*) (J_a^* + J_b^* + 2 \Lambda^*) } \right]  \\
 = & ~ \text{Re} \: \left[ \frac{(1 - a g) (b + i r_+) }{3 g r_+^2 - 2 i (1 + a g + b g) r_+ - (a + b + a b g)} ( \varphi g + 2 \pi i T ) \right]  \\
 = & ~ - \frac{g^3 (1 - a g) (1 + a g + b g) (3 a + b + a b g + b^2 g) B}{4 \pi^2 a^2 b (a + b)^2 (1 + a g) (1 + b g)^2 }  \varphi \nn\\
& ~ - \frac{(1 -a g) \left( b - a + (3 b^2 + 4 a b - a^2) g + b (2 b^2 + 3 a b + a^2) g^2 \right)g^2  B}{2\pi  a b r^* (a + b)^2 (1 + a g)^2 (1 + b g)^2} T,
\end{align}
and
\begin{align}
\text{Re}  \: (\Omega_b - \Omega_b^*) = &~\text{Re} \: \left[\frac{  (J_a^* + \Lambda^* )( 3 g \Lambda^*- Q^*) \left( \varphi g + 2 \pi i T\right)}
{9 g (J_a^* + \Lambda^*) (J_b^* + \Lambda^*) - ( 3 g \Lambda^*- Q^*) (J_a^* + J_b^* + 2 \Lambda^*) } \right] \\
 =&~\text{Re} \: \left[ \frac{(1 - b g) (a + i r_+) }{3 g r_+^2 - 2 i (1 + a g + b g) r_+ - (a + b + a b g)} ( \varphi g + 2 \pi i T ) \right]  \\
 =&~ - \frac{g^3 (1 - b g) (1 + a g + b g) (3 b + a + a b g + a^2 g) B}{4 \pi^2 a b^2 (a + b)^2 (1 + a g)^2 (1 + b g) }  \varphi \nn\\
&~ - \frac{(1 -b g) \left( a - b + (3 a^2 + 4 a b - b^2) g + a (2 a^2 + 3 a b + b^2) g^2 \right)g^2  B}{2\pi  a b r^* (a + b)^2 (1 + a g)^2 (1 + b g)^2} T.
\end{align}
Similarly, for  the electric potential, we find
\begin{align}
\text{Re} \:  (\Phi - \Phi^*) =&~\text{Re} \: \left[\frac{3  (J_a^* + \Lambda^* )(J_b^* + \Lambda^* )\left( \varphi g + 2 \pi i T\right)}
{9 g (J_a^* + \Lambda^*) (J_b^* + \Lambda^*) - ( 3 g \Lambda^*- Q^*) (J_a^* + J_b^* + 2 \Lambda^*) } \right] \\
 =&~\text{Re} \: \left[ \frac{ (r_+ - i a) (r_+ - i b)}{3 g r_+^2 - 2 i (1 + a g + b g) r_+ - (a + b + a b g)} ( \varphi g + 2 \pi i T ) \right]  \\
=&~ \frac{g^3 (2 + 5(a+b)g +(3a^2 +4 a b + 3 b^2 ) g^2 + a b (a + b) g^3) B}{4 \pi^2 a b (a + b) (1 + a g)^2 (1 + b g)^2 }  \varphi 
\nn \\
&~ 
- \frac{ \left( a + b + ( a-b)^2 g -  a b(a+b )g^2 \right)  g^2 B}{2\pi  a b r^* (a + b) (1 + a g)^2 (1 + b g)^2} T.
\end{align}

Constraint \eqref{new_cc} on the complex potentials gives
\begin{align}
3 g \; \text{Re} \; (\Phi - \Phi^*) -\text{Re}\;  (\Omega_a - \Omega_a^*) - \text{Re}\;  (\Omega_b - \Omega_b^*)  = \varphi g .
\end{align}
Identifying the gravitational potentials with the real parts, we obtain the following relations for near-BPS black holes,
\begin{align}
3 g  (\Phi^{\mathrm{bh}} - \Phi^{\mathrm{bh}}{}^*) -  (\Omega_a^{\mathrm{bh}} - \Omega_a^{\mathrm{bh}}{}^*) - (\Omega_b^{\mathrm{bh}} - \Omega_b^{\mathrm{bh}}{}^*)  = \varphi g.
\end{align}
In order to avoid any confusion we use the superscript $^\mathrm{bh}$ to denote the black hole quantities, as opposed to the complex quantities that enter the extremization principle. Taking $\varphi$ and $T$ derivatives respectively, we get
\begin{align}
3 \partial_\varphi \Phi ^{\mathrm{bh}} &= \frac{1}{g} \left(\partial_\varphi \Omega_a^{\mathrm{bh}} + \partial_\varphi \Omega_b^{\mathrm{bh}} \right) + 1, &
3 \partial_T \Phi ^{\mathrm{bh}}&= \frac{1}{g} \left( \partial_T \Omega_a^{\mathrm{bh}} + \partial_T \Omega_b^{\mathrm{bh}} \right).
\end{align}
It can be readily checked that all these relations are satisfied. All these expressions also agree with the analysis of subsections 
\ref{sec:near_BPS_BH} and 
\ref{sec:complex_potentials}. 
 In fact, subsection \ref{sec:complex_potentials} contains very much the same information as the present subsection; only the organisation is 
slightly different.

\section{Phase diagram of BPS black holes}
\label{sec:phase_diagram}
In this section we first introduce appropriate notions of the BPS free energy and BPS temperature. Then we study the phase diagram of BPS black holes in six-dimensions. Our presentation follows \cite{Ezroura:2021vrt}.

\subsection{BPS thermodynamics}
The Gibbs free energy $\mathbb{G}(T, \Phi, \Omega_a, \Omega_b), $
\begin{align}\label{fe}
\mathbb{G} = M - T S - \Phi Q - \Omega_a J_a - \Omega_b J_b,
\end{align}
for the BPS black holes is identically zero. Hence, it is not a useful quantity to work with. 
Instead, it is useful to define the BPS free energy \cite{Ezroura:2021vrt},
\begin{align}
W (\Phi{'} , \Omega_a{'} , \Omega_b{'} ) = \frac{\mathbb{G}}{T}  = -  S^* - \Phi{'} Q^* -\Omega_a{'}  J_a^* - \Omega_b{'}  J_b^*,
\end{align}
where the primed potentials are defined as,
\begin{align}
& \Phi{'} = \frac{(\Phi - \Phi^*)}{T}, \\  & \Omega_a{'} = \frac{(\Omega_a - \Omega_a^*) }{T} , \\  & \Omega_b{'} = \frac{(\Omega_b - \Omega_b^*) }{T},
\end{align}
and the BPS limit is understood.

To begin with, we would like to compute $W (\Phi{'} , \Omega_a{'} , \Omega_b{'} )$ for the AdS$_6$ BPS black holes. We can do this as follows: as discussed in detail in section \ref{sec:gravity} the BPS limit corresponds to $r_+ = r^*$ and $q = q^*$. Therefore, quantity \eqref{fe}  written in terms of the black hole charges  \eqref{BPSM}--\eqref{BPSJb} and potentials \eqref{cp}--\eqref{OmegaB} has an expansion in powers of $ r_+ - r^*$ and $q - q^*$. We start by converting such an expansion  in terms of $\Phi - \Phi^*$, $\Omega_a - \Omega_a^*$, and $\Omega_b - \Omega_b^* $. Then upon dividing by $T$ we obtain $W$ as a function of  $\Phi{'} $, $\Omega_a{'}$, $\Omega_b{'}$.

Using equation (\ref{exp_phi}) we can write to order\footnote{In this subsection, to keep the notation simple we do not write ${\mathcal O}(\epsilon^2)$ in most of the equations.}  $\epsilon$
\be
 q - q^* = \frac{\Phi - \Phi^*}{\Phi_q} - \frac{(r_+ - r^*) \Phi_r}{\Phi_q} + \mathcal{O}(\epsilon^2). \label{delta_q}
\ee  
Substituting equation \eqref{delta_q} 
 in equation (\ref{exp_om}) we obtain, 
\begin{align}\label{ex_o_r_a}
\Omega_a - \Omega_a^* &=- \frac{ 2 b g (1-a g ) }{r^* (a+b) (1+ a g)} (r_+ - r^*)  - \frac{(1-a g)}{a+b} (\Phi - \Phi^*), \\
\Omega_b - \Omega_b^* &=- \frac{ 2 a g (1- b g) }{r^* (a+b) (1+ b g)} (r_+ - r^*)  - \frac{(1-b g )}{a+b} (\Phi - \Phi^*). 
\label{ex_o_r_b}
\end{align}
Solving for  $ r_+ - r^*$ from equations \eqref{ex_o_r_a} and \eqref{ex_o_r_b} we get, 
\begin{align}
r_+ -r^*  = &  - \frac{r^*}{4 g a b } \left(
   \frac{ a (a + b) (1+ a g )}{(1- a g)} ( \Omega_a - \Omega_a^*)  + \frac{
    b (a + b) (1+ b g)}{(1-b g )}   ( \Omega_b - \Omega_b^*) \right) \nn  \\ 
    &    - \frac{r^*}{4 g a b }   \left( a ( 1+ ag ) + b ( 1+ bg) \right)
     ( \Phi - \Phi^*). \label{delta_r}
\end{align}
Equations \eqref{delta_r} and \eqref{delta_q} provide the substitutions for $r_+ - r^*$ and $q - q^*$ that we need for finding $W$.
Moreover, the condition that the two equations \eqref{ex_o_r_a} and \eqref{ex_o_r_b}  give identical values for $r_+-r^*$ yields a constraint on the potentials, 
\begin{align} \label{con_1}
\left( \frac{ 1+ a g}{1-a g} \right) \frac{\Omega_a -\Omega_a^*}{b}  - \left( \frac{ 1+ b g}{1-b g} \right) \frac{\Omega_b -\Omega_b^*}{a}    + \left( \frac{a - b}{a+b} \right) \frac{\Phi-\Phi^*}{r^{*2}}= 0.
\end{align}
Thus, we have managed to write $ r_+ - r^*$ and $q - q^*$ in terms of $\Phi - \Phi^*$, $\Omega_a - \Omega_a^*$, and $\Omega_b - \Omega_b^* $, provided constraint \eqref{con_1} is satisfied.

Using these substitutions, temperature \eqref{temp2}  can be written in terms of  $\Phi-\Phi^*$, $\Omega_{a}-\Omega_{a}^* $, and  $\Omega_{b}-\Omega_{b}^* $ as
\begin{align} \label{con_2_temp}
T =
     &  - \left( \frac{4 (a + b) + (7 a^2 + 6 a b + 7 b^2) g + (a + b) (3 a^2 + 2 a b + 
    3 b^2) g^2 + a b (a + b)^2 g^3 }{4 \pi (a + b)  (1 + a g) (1 + b g) r^*}\right)  ( \Phi - \Phi^*) \nn \\
&- r^* \left(\frac{2 +5  (a+b)g +   \left(3 a^2+4 a b+3 b^2\right)g^2+a b 
   (a+b)g^3 }{4 \pi b (1 - a g) (1 + b g)} \right) \left( \Omega_a - \Omega_a^* \right) \nn \\
&- r^* \left(\frac{2 +5  (a+b)g +   \left(3 a^2+4 a b+3 b^2\right)g^2+a b 
   (a+b)g^3 }{4 \pi a (1 - b g) (1 + a g) } \right) \left( \Omega_b - \Omega_b^* \right).
   \end{align}

The Gibbs free energy $\mathbb{G}$ written as an expansion to first order in  $\Phi-\Phi^*$, $\Omega_{a}-\Omega_{a}^* $, and  $\Omega_{b}-\Omega_{b}^* $ takes the form,
\begin{align}\label{exp_G}
\mathbb{G} = 
&  -\frac{\pi (a+ b)  (2 + (a+b)g - (a+b)^2 g^2)r^{*3}}{6 a b g G (1-ag)(1-bg)  }(\Phi - \Phi^*) +
 \frac{\pi (a+ b)^2  (1+ag)r^{*3}}{6 b  G (1-ag)^2(1-bg)  }(\Omega_a -\Omega_a^*)   \nn \\ 
 &+ \frac{\pi (a+ b)^2  (1+bg)r^{*3}}{6 a  G (1-ag)(1-bg)^2  }(\Omega_b -\Omega_b^*).
\end{align}
The BPS free energy $W$  is obtained from \eqref{exp_G}  by dividing with $T$. We obtain our final expression: 
\begin{align} \label{W_final}
W = 
&  -\frac{\pi (a+ b)  (2 + (a+b)g - (a+b)^2 g^2)r^{*3}}{6 a b g G (1-ag)(1-bg)  }\Phi{'}
+ \frac{\pi (a+ b)^2  (1+ag)r^{*3}}{6 b  G (1-ag)^2(1-bg)  }\Omega_a{'}    \nn \\ & + \frac{\pi (a+ b)^2  (1+bg)r^{*3}}{6 a  G (1-ag)(1-bg)^2  } \Omega_b{'}.
\end{align}

In terms of primed variables constraint \eqref{con_1} takes the form,
\begin{align} \label{con_1_1}
\left( \frac{ 1+ a g}{1-a g} \right) \frac{\Omega_a'}{b}  - \left( \frac{ 1+ b g}{1-b g} \right) \frac{\Omega_b'}{a}    + \left( \frac{a - b}{a+b} \right) \frac{\Phi'}{r^{*2}}= 0.
\end{align}
Expression \eqref{con_2_temp}  for temperature  becomes another constraint,
\begin{align} \label{con_2}   
1= &  - \left( \frac{4 (a + b) + (7 a^2 + 6 a b + 7 b^2) g + (a + b) (3 a^2 + 2 a b + 
    3 b^2) g^2 + a b (a + b)^2 g^3 }{4 \pi (a + b)  (1 + a g) (1 + b g) r^*}\right)  \Phi' \nn \\
&- r^* \left(\frac{2 +5  (a+b)g +   \left(3 a^2+4 a b+3 b^2\right)g^2+a b 
   (a+b)g^3 }{4 \pi b (1 - a g) (1 + b g)} \right) \Omega_a'  \nn \\
&- r^* \left(\frac{2 +5  (a+b)g +   \left(3 a^2+4 a b+3 b^2\right)g^2+a b 
   (a+b)g^3 }{4 \pi a (1 - b g) (1 + a g) } \right)  \Omega_b'.
   \end{align}
In the grand canonical ensemble, parameters $a,b$ are not independent. They are complicated functions of $\Phi'$ and $\Omega_{a}'$ and $\Omega_{b}'$. In principle, they are obtained by solving the two constraints \eqref{con_1_1} and \eqref{con_2}.

\subsection{BPS free energy}

In terms of variable $\varphi$ introduced in \eqref{vphi}, we have upon taking the $T$ derivative, 
\begin{align}
\varphi' =  3 \Phi'  - g^{-1}  ( \Omega_a'   + \Omega_b').
\end{align}
We can eliminate $\Phi'$ in favour of $\varphi' $, $\Omega_a' $, $\Omega_b'$. Substituting for $\Phi' $ in equations \eqref{con_1_1} and \eqref{con_2} we get two equations for $\Omega_a' $, $\Omega_b'$. These equations can be solved to give $\Omega_a' $, $\Omega_b'$ in terms of $a, b, \varphi'$. We have,
\begin{align}\label{Omega_a_prime}
\Omega_a' = & \frac{2 \pi r^{*3}  }{A_2 a b } ( a - b - 3 b (a + b) g - (a^3 - 3 a^2 b + 2 b^3) g^2 + 
  a b (a^2 + 3 a b + 2 b^2) g^3)
\nn \\ & - \left[ \frac{r^{*2}  \Xi_a }{A_2 a } g ( 3 a + b +   (a + b)b g) \right] \varphi',
\end{align}
and
\begin{align}\label{Omega_b_prime}
\Omega_b' = & \frac{2 \pi r^{*3}  }{A_2 a b } ( b - a - 3 a (a + b) g - (b^3 - 3 b^2 a + 2 a^3) g^2 + 
  a b (a^2 + 3 a b + 2 b^2) g^3) \nn  \\ & - \left[ \frac{r^{*2}  \Xi_b }{A_2 b } g ( 3 b + a +  (a + b) a g) \right] \varphi', 
\end{align}
where $A_2$ was introduced in \eqref{A_2}. 
With various variables introduced above it is easy to check (using Mathematica) that
\be
 \Omega_i' \left(a, b, \varphi' = \frac{2 \pi i}{g} \right) = \omega_i (a,b),
\ee
where $\omega_i (a,b)$ are given in equations \eqref{omega_a_BPS_choi}--\eqref{omega_b_BPS_choi}. Similarly,
\be
\Phi'\left(a, b, \varphi' = \frac{2 \pi i}{g} \right) = \Delta (a,b),
\ee
where $\Delta (a,b)$ are given in \eqref{Delta_BPS_choi}. These equations are a rewriting of the observations already made  in section \ref{sec:complex_potentials}.

Substituting relations \eqref{Omega_a_prime}--\eqref{Omega_b_prime}  in equation \eqref{W_final} we obtain the BPS free energy as,
\begin{align}\label{W_final_varphi}
&W = W_1 + W_2 \varphi',\\ 
\label{W_final_varphi1}
&W_1 = \frac{2 \pi ^2 r^{*6} (a + b)^2}{(3 A_2  a^2 b^2 g (1 - a g) (1 - b g) G))} 
\Big{(}a + b + (2 a^2 - 2 a b + 2 b^2) g + (a^3 - 7 a^2 b - 7 a b^2 + 
      b^3) g^2 \nn
      \\  & \qquad \qquad - (4 a^3 b + 6 a^2 b^2 + 4 a b^3) g^3 - (a^3 b^2 + 
      a^2 b^3) g^4)\Big{)}, \\
\label{W_final_varphi2}
&W_2     = -\frac{2 \pi  r^{*5} (a + b)^2}{3 A_2  a b (1 - a g) (1 - b g)G} (1 + 3 (a + b) g + (2 a^2 + a b + 2 b^2) g^2).
\end{align}
We can relate this free energy to  the on-shell action. We have \cite{Cassani:2019mms},  
\be
W \left(a, b,\varphi' = \frac{2 \pi i}{g} \right) = I^*(a,b),
\ee
where $I^*(a,b)$ is the on-shell action \eqref{action} for the BPS black holes. We also note that, 
\be
W \left(a, b,\varphi' = 0 \right) = \mathrm{Re} \: I^*(a,b). 
\ee

Expressions \eqref{W_final_varphi}--\eqref{W_final_varphi2} are the main results of this subsection. We use these expressions in the following subsections to study the phase diagrams of BPS black holes.

\subsection{Phase diagram for $\varphi' =0$} \label{4.3}

\begin{figure}[t!]
\begin{center}
  \includegraphics[width=.6\textwidth]{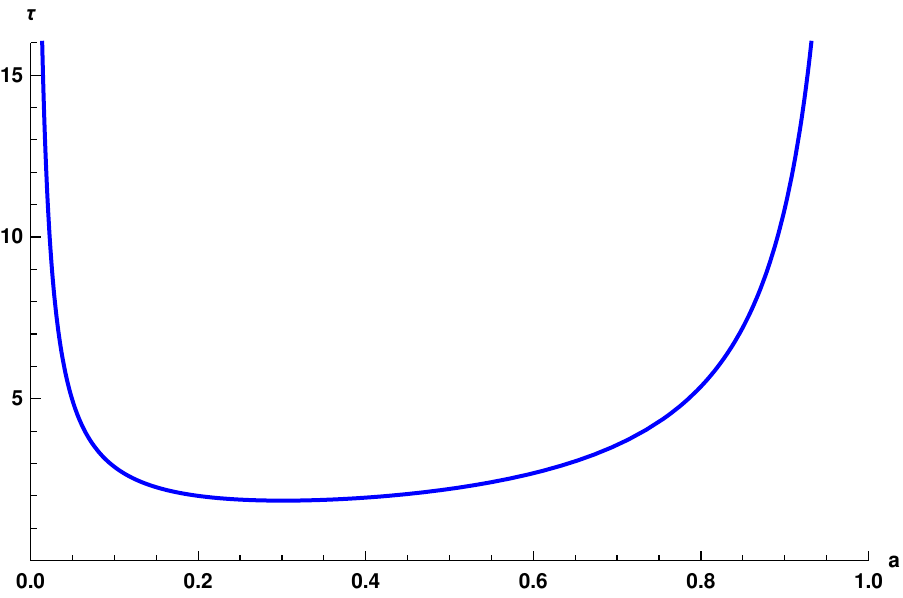}
   \caption{ The BPS temperature $\tau$ as a function of the rotation parameter $a$, when $a=b$ and $\varphi'=0$. The large-$\tau$ extremes of the small and large black hole branches are shown by the divergences at $a=0$ and $a=1$ respectively.}
   \label{fig:1}
   \end{center}
\end{figure}

The phase diagram of five-dimensional BPS black holes has attracted a lot of interest in the recent years \cite{Choi:2018vbz, Choi:2018hmj, Copetti:2020dil, Choi:2021lbk, Ezroura:2021vrt}.  Motivated by these developments, we find it  useful to study the phase diagram for AdS$_6$ BPS black holes. We only focus on the equal angular momentum cases.  To the best of our knowledge, a microscopic study of the type \cite{Choi:2018vbz, Choi:2018hmj} has not been performed for the 5d indices relevant for the 6d black holes.

Although the temperature of the BPS black holes is zero, it is useful to introduce a BPS temperature \cite{Choi:2018vbz, Choi:2018hmj, Ezroura:2021vrt},
\be \label{tau_varphi}
\tau = - \frac{2 g}{ \Omega_a' +  \Omega_b'}.
\ee
We use $(a , b, \varphi')$ as variables for all thermodynamic quantities. The BPS temperature simplifies when  $a = b$ and $\varphi'=0$ (in this and the next subsection we set $g=1$ and $G=1$):
\begin{align}\label{btemp}
\tau &= \frac{2 + 6 a + 9 a^2 + a^3 }{3 \pi  (1-a) a \sqrt{1+ 2a}}.
\end{align}

In the range $0 < a < 1$ the BPS temperature is positive and finite; it diverges at $a=0$ and $a=1$.   Function $\tau(a)$ has a local minima at $a_\mathrm{cusp} \approx  0.301$  with value $\tau_\mathrm{cusp} \approx 1.852$.
It is plotted in Fig.~\ref{fig:1}.

The BPS free energy \eqref{W_final_varphi} when $a = b$ and $\varphi'=0$ simplifies to 
 \be \label{Wn}
W = \frac{4 \pi ^2 a^3 \left(1-6 a^2-a^3\right)}{3
   (1-a)^2 (1+ 2 a) \left(2 + 6 a + 9 a^2 + a^3\right)}.
\ee

\begin{figure}[h!]
\begin{center}
 \includegraphics[width=0.8\textwidth]{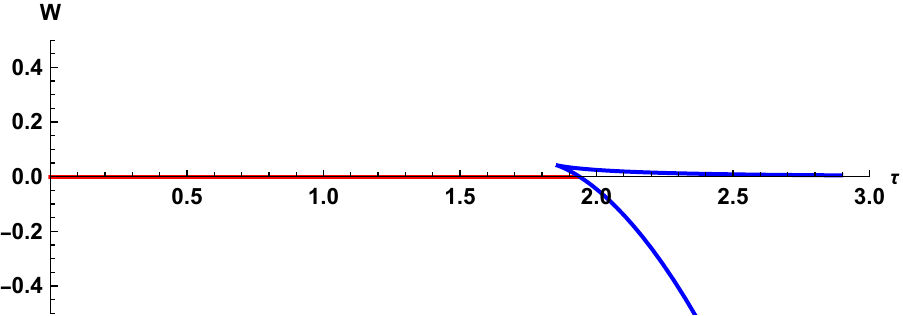}
   \caption{ The BPS free energy $W$ vs. the BPS temperature $\tau$ as a function of the rotation parameter $a$, when $a=b$ and $\varphi'=0$. The upper branch of the phase diagram is referred to as the ``small'' black hole branch and the lower branch is  referred to as ``large'' black hole branch. The two branches meet at the cusp $(\tau_\mathrm{min} , W_\mathrm{max})  \approx (1.852 , 0.0423)$. The red line $W=0$ in the range $0< \tau < \tau_\mathrm{HP}$ represents the ``thermal BPS'' gas. The large black holes are thermodynamically preferred for $\tau > \tau_\mathrm{HP}  \approx 1.937$. }
   \label{fig:2}
   \end{center}
\end{figure}

\begin{figure}[h!]
\begin{center}
 \includegraphics[width=0.8\textwidth]{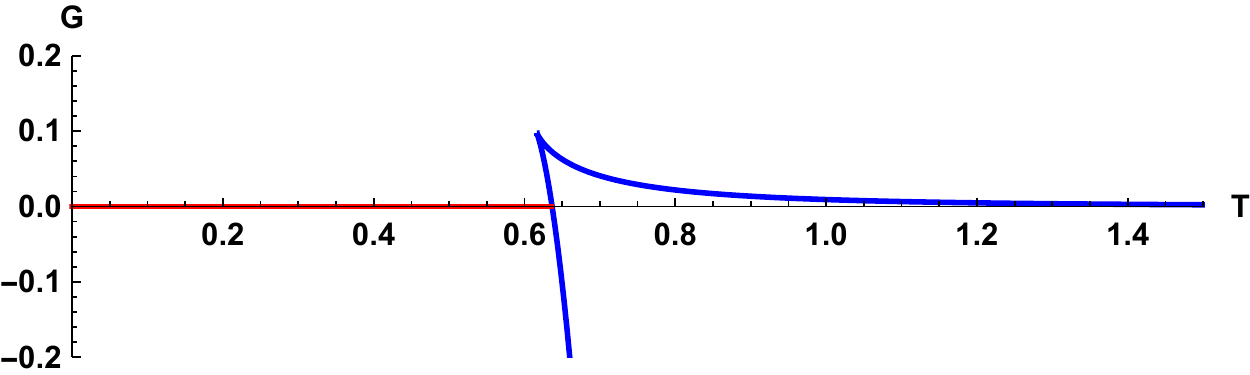}
   \caption{ The AdS-Schwarzschild phase diagram. The Gibbs free energy $G$ vs. the temperature $T$. The upper branch of the phase diagram is referred to as the small black hole branch and the lower branch is referred to as the large black hole branch. The two branches meet at the cusp  $(T_\mathrm{min} , G_\mathrm{max})  \approx (0.616 , 0.097)$. The red line $G=0$ in the range $0< T < T_\mathrm{HP}$ represents the thermal AdS gas. The large black holes are thermodynamically preferred for $T  > T_\mathrm{HP}  \approx 0.636$. }
   \label{fig:3}
   \end{center}
\end{figure}

The BPS phase diagram $W$ vs.~$\tau$  is shown in Figure \ref{fig:2}. Sailent features are:
\begin{itemize}
\item[$\bullet$]  The BPS free energy $W$ goes to zero and the BPS temperature $\tau$ diverges as the parameter  $a$ approaches  zero. 
\item[$\bullet$]   As $a$ increases from zero,   the BPS temperature starts decreasing and reaches its minimum value  $\tau_\mathrm{cusp} \approx 1.852 $ at $a_\mathrm{cusp} \approx 0.301$, while the BPS free energy increases monotonically to its maximum value $W_\mathrm{cusp} \approx 0.0423$.
\item[$\bullet$]  In the range $0 < a < a_\mathrm{cusp} $,  the BPS free energy is positive and  maps out the \enquote{small} black hole branch (the upper branch)  of the BPS phase diagram \ref{fig:2}.  In this region, the small black hole phase is thermodynamically unstable. 
\item[$\bullet$]     The range $a_\mathrm{cusp} < a <1$ maps out the \enquote{large} black hole branch (the lower branch) of the BPS phase diagram Figure \ref{fig:2}. On this branch, the free energy decreases from $W_\mathrm{cusp}$ and become $W=0$ at $a_\mathrm{HP} \approx  0.3954$. The value $a_\mathrm{HP}$ is the analog of the  Hawking-Page transition point for the AdS-Schwarzschild black holes. The Hawking-Page temperature is $\tau_\mathrm{HP} \approx 1.937$.   For $\tau < \tau_\mathrm{HP}$ the ``thermal BPS gas phase'' (shown as red line in Figure \ref{fig:2})  dominates over AdS black hole phases.\footnote{The notion of ``thermal BPS gas'' has not yet been made precise. See comments in \cite{Ezroura:2021vrt}.} 
\item[$\bullet$]  In the range $a_\mathrm{HP} < a <1$,  the BPS free energy is negative.  In this region, the large black hole phase is  thermodynamically preferred. 
\item[$\bullet$]  Finally, in the limit $a \rightarrow 1^- ,$  the BPS free energy diverges $W \rightarrow  - \infty$  along with the BPS temperature $\tau \rightarrow  \infty$.
\end{itemize}

It is instructive to compare the BPS phase diagram with the the six-dimensional AdS-Schwarzschild phase diagram (for a comprehensive review of AdS black hole thermodynamics, see \cite{Kubiznak:2016qmn}). The Gibbs free energy \eqref{fe}  and the temperature \eqref{temp} for the  AdS-Schwarzschild take the form:
\begin{align}
\mathbb{G} &= \frac{1}{6} \pi r_+^3 (1 - r_+^2), &
T &= \frac{3 + 5 r_+^2 }{4 \pi r_+}, & W & := \frac{\mathbb{G}}{T} = \frac{2 \pi^2 r_+^4 (1 - r_+^2)}{9 + 15 r_+^2}. \label{Wfes}
\end{align}
The AdS-Schwarzschild phase diagram is shown in Figure \ref{fig:3}.

Although, the phase diagram of AdS-BPS black holes  Fig.~\ref{fig:2} appears qualitatively similar to that of the AdS-Schwarzschild  black holes Fig.~\ref{fig:3}, there are several differences.  

On the one hand, for very small AdS-Schwarzschild black holes $ r_+ \ll 1$, $T$ diverges as $r_+^{-1}$ and $W$ vanishes as $r_+^4$ so:
\be 
W \sim T^{-4} \quad \mathrm{as} \quad T \rightarrow \infty. 
\ee
  For very large black holes $ r_+ \gg 1$, $T$ diverges as $r_+$ and $W$ diverges as $- r_+^4$ so:
\be
 W \sim - T^{4} \quad \mathrm{as} \quad T \rightarrow \infty.
 \ee

On the other hand, for very small BPS black hole $a \rightarrow 0^+$, $\tau$ diverges as $a^{-1}$ and $W$ vanishes as $a^3$ so: 
\be
 W \sim \tau^{-3} \quad \mathrm{as} \quad \tau \rightarrow \infty.
 \ee
 For large BPS black hole $a \rightarrow 1^-$, $\tau$ diverges as $(1-a)^{-1}$ and $W$ diverges as $- (1-a)^{-2}$ so: 
\be
 W \sim - \tau^{2} \quad \mathrm{as} \quad \tau \rightarrow \infty.
 \ee 
 Clearly, there are differences.

\subsection{Phase diagram for $\varphi' \neq 0$}
In this final subsection we turn on the the potential $\varphi'$. This potential qualitatively modifies the phase diagram. It is important to appreciate that potential $\varphi'$ is not captured by the supersymmetric index, but turning it on preserves the BPS saturation.

 For simplicity, we only consider  the case of  equal angular momenta. The BPS free energy (\ref{W_final_varphi})  simplifies to,  
\be \label{Wn2}
W = \frac{4 \pi ^2 a^3 \left(1-6 a^2-a^3\right)}{3
   (1-a)^2 (1+ 2 a) \left(2 + 6 a + 9 a^2 + a^3\right)} 
   -   \frac{2\pi  a^3 \left(1+5a\right)}{3
   (1-a)^2 \sqrt{1+ 2 a} \left(2 + 6 a + 9 a^2 + a^3\right)} \varphi' .
\ee
The BPS temperature $\tau$ (\ref{tau_varphi}) simplifies to
\begin{align}\label{btemp2}
\tau &= \frac{2(2 + 6 a + 9 a^2 + a^3) }{  (1-a) \left( 6 \pi a \sqrt{1+ 2a} + (1 + 2 a) (2 + a ) \varphi' \right)} .
\end{align}
In the following two sections we study the phase diagrams for different signs of $\varphi'$.

 \subsubsection{$\varphi' < 0$}
In the previous subsection we saw that the physical range of the BPS black holes is $0 < a < 1$ for $\varphi' =0$. At the two ends, the BPS temperature diverges. For $\varphi' \neq 0$, the BPS temperature \eqref{btemp2} diverges as  $a \to 1$. For $\varphi' < 0$, the physical range of the BPS black holes shrinks to $a_\mathrm{min} <a <1$, as the second factor in the denominator of equation (\ref{btemp2}) vanishes at $a_\mathrm{min}$ where,
 \begin{align}
6 \pi a_\mathrm{min} +  \sqrt{1+ 2 a_\mathrm{min}} (a_\mathrm{min} + 2) \varphi'   = 0.
 \end{align}
For small negative values of $\varphi'$ we see that $a_\mathrm{min} \approx - \frac{\varphi'   }{3 \pi}$.  $a_\mathrm{min}$ increases with decreasing $\varphi'$. Potential $\varphi' $ reaches its lower bound as $a_\mathrm{min}$ approaches 1, where
 \begin{align}
\varphi'_- = \left. \varphi' \right\vert_{a_\mathrm{min} =1} = - \frac{2 \pi}{\sqrt{3}}.
 \end{align}
 In the strict limit $\varphi' \to \varphi'_-$, no physical black holes remain.
 
 For $\varphi' < 0$, the BPS phase diagram  qualitatively changes compared to the $\varphi' =0$ case: the asymptotic value of the BPS free energy is non-zero on the small black hole branch. As parameter $a$ approaches $a_\mathrm{min} (\varphi' )$, the BPS free energy approaches a finite non-zero value:
\begin{align}
W_\mathrm{asym} (a_\mathrm{min} (\varphi')) = \frac{4 \pi ^2 a_\mathrm{min}^3}{ \left(6 + 9 a_\mathrm{min} -9 a_\mathrm{min}^2 - 6 a_\mathrm{min}^3\right)}.
\end{align}
As $\varphi' \to 0$, $a_\mathrm{min} \to 0$ and  $W_\mathrm{asym} \to 0$, consistent with the result of the previous subsection.

The $a \to 1^-$ limit gives the asymptotic relation between the BPS free energy and the BPS temperature on the large black hole branch. We have,
\begin{align}
W \simeq -  \frac{\pi \left(2 \pi + \sqrt{3} \varphi' \right)^3}{648} \tau^2. 
\end{align} 

 \begin{figure}[t]
\begin{center}
 \includegraphics[width=0.9\textwidth]{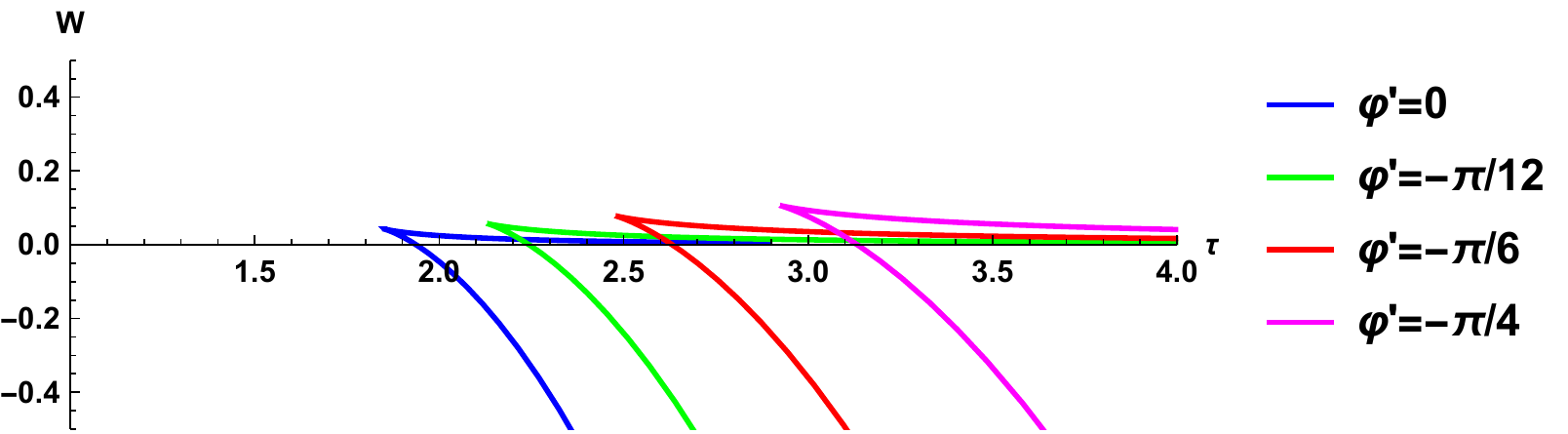}
   \caption{ The BPS free energy $W$ as function of the BPS temperature $\tau$ for negative values of $\varphi'$. The two angular momenta are equal $(a = b)$ and $\varphi' = 0, - \pi/12, - \pi/6 , - \pi/4,$ from left to right. For each value of $\varphi'$, there are two branches  that meet in a cusp. The small black hole banch (the upper branch) asymptote to a positive BPS free energy at large temperature for $\varphi' < 0$. }   
   \label{plot:1}
   \end{center}
\end{figure}

The minimum BPS temperature for given $\varphi'$ is attained at $a_\mathrm{cusp}$ determined by $ \partial_a \tau =0$. This leads to the following non-linear equation, 
\begin{align} \label{min_temp}
\varphi'=\frac{2 \pi 
\left(2 + 2 a_\mathrm{cusp} - 19 a_\mathrm{cusp}^2 - 29 a_\mathrm{cusp}^3 - 11 a_\mathrm{cusp}^4 + a_\mathrm{cusp}^5 \right)}{\sqrt{1+ 2 a_\mathrm{cusp}} 
\left(2 + 16 a_\mathrm{cusp} + 21 a_\mathrm{cusp}^2 + 10 a_\mathrm{cusp}^3 + 5 a_\mathrm{cusp}^4 \right)}.
\end{align}
In the limit $\varphi' \rightarrow  0$, this equation can be solved to yield 
$ a _\mathrm{cusp} \approx 0.301$,  which agrees with the result of the previous subsection. As  $\varphi'$ decreases from 0 to $\varphi' _-$,  $a_\mathrm{cusp} $  monotonically increases  from $ a _\mathrm{cusp} \approx 0.301$ to $ a _\mathrm{cusp} = 1$.

To explore the effect of  $\varphi'$ on the stability of the BPS black holes, we study the changes in  the value of the Hawking-Page temperature and the BPS free energy  at the cusp.

The Hawking-Page temperature corresponds to $W=0$. For a given $\varphi'$, $W=0$ at $a=a_\mathrm{HP}$, given of the solution to the equation,
\begin{align}
\varphi' = \frac{2 \pi \left(1-6 a_\mathrm{HP}^2-a_\mathrm{HP}^3\right)}{  \sqrt{1+ 2 a_\mathrm{HP}}(1 +5 
  a_\mathrm{HP})}.
\end{align}
The derivative of this expression is negative in the full range $0\le a_\mathrm{HP} \le 1$,
\be
\partial_{a_\mathrm{HP}} \varphi' = -\frac{6 \pi \left(1 + a_\mathrm{HP} \right)^2 \left(2 + 5 a_\mathrm{HP} + 5 a_\mathrm{HP}^2\right)}{(1 + 2 a_\mathrm{HP})^{\frac{3}{2}} \left(1 + 5 a_\mathrm{HP}\right)^2}.
\ee
Therefore, $a_\mathrm{HP}$ increases monotonically as $\varphi'$ decreases. 
The change in the Hawking-Page temperature with changing $\varphi'$ can be conveniently capture by the $\frac{d}{d\varphi'}$ derivative,
\begin{align}
  \frac{d}{d \varphi' } \tau_\mathrm{HP} (\varphi', a_\mathrm{HP} (\varphi') ) &= (\partial_{\varphi'}  \tau) \bigg{|}_{a = a_\mathrm{HP}} + (\partial_{a}  \tau) \bigg{|}_{a = a_\mathrm{HP}} \; \partial_{\varphi'} a_\mathrm{HP}  \\ 
  &= (\partial_{\varphi'}  \tau) \bigg{|}_{a = a_\mathrm{HP}} + (\partial_{a}  \tau) \bigg{|}_{a = a_\mathrm{HP}} \; \left( \partial_{a_\mathrm{HP}} \varphi' \right)^{-1} \\ 
  & = - \frac{(1+ 5 a_\mathrm{HP} )^2}{2 \pi^2  (1-a_\mathrm{HP} )^3 (1+ a_\mathrm{HP} )^2}   < 0.
\end{align}
Thus, the Hawking-Page temperature temperature is a \emph{decreasing} function of $\varphi'$. As $\varphi'$ decreases, $\tau_\mathrm{HP}$ increases. This feature is clearly seen in Fig.~\ref{plot:1}.

We can analogously study the  changes in the value of the BPS energy $W$ at the cusp with changing $\varphi'$.  We have,
\begin{align}
  \frac{d}{d \varphi' }W_\mathrm{cusp}  (\varphi', a_\mathrm{cusp} (\varphi') ) &= (\partial_{\varphi'}  W) \bigg{|}_{a = a_\mathrm{cusp}} + (\partial_{a}  W) \bigg{|}_{a = a_\mathrm{cusp}} \; \partial_{\varphi'} a_\mathrm{cusp}  \\ 
  &= (\partial_{\varphi'}  W) \bigg{|}_{a = a_\mathrm{cusp}} + (\partial_{a}  W) \bigg{|}_{a = a_\mathrm{cusp}} \; \left( \partial_{a_\mathrm{cusp}} \varphi' \right)^{-1} \\ 
  & = - \frac{2 \pi a_\mathrm{cusp}^3 (1+ 5 a_\mathrm{cusp} )}{3 (1-a_\mathrm{cusp} )^2 \sqrt{1 + 2 a_\mathrm{cusp}}( 2+ 6 a_\mathrm{cusp} + 9 a_\mathrm{cusp}^2 + a_\mathrm{cusp}^3)}   < 0.
\end{align}
 Thus, as $\varphi'$  decreases from $0$ to $\varphi' _-$, the cusp value of the BPS free energy increases monotonically. This feature is also clearly seen in Fig.~\ref{plot:1}. To summarise: Compared to the $\varphi' =0$ case discussed in the previous subsection,
\begin{itemize}
\item[$\bullet$] The BPS free energy $W_\mathrm{cusp}$ at the cusp  increases with decreasing $\varphi'$ below zero.
\item[$\bullet$] The Hawking-Page temperature  $\tau_\mathrm{HP}$ increases with decreasing $\varphi'$ below zero.
\end{itemize}   
These features indicate that  decreasing $\varphi'$ below zero thermodynamically destabilises the black hole.

\subsubsection{$\varphi' > 0$}

Unlike $\varphi' <  0$,  when $\varphi' > 0$ the denominator of the BPS temperature \eqref{btemp2} only diverges as  $a \to 1$. The second term in the denominator is strictly positive for $a \in [0,1)$. The entire range $ 0 < a < 1$ is thus physical.  As   $a \rightarrow 0$, the BPS free energy vanishes and the BPS temperature $\tau$ reaches its maximum value on the small black hole branch, 
 \begin{align}\label{t_max}
\tau_\mathrm{max} = \frac{2}{\varphi'}.
\end{align}
In the limit $a \to 0$, the BPS black hole geometry degenerates to pure AdS$_6$; it is not a black hole. In this sense, $\tau_\mathrm{max}$ is a bound, it cannot be reached. The range of the allowed BPS temperatures on the small black hole branch is lowered as $\varphi'$ increases. In the limit $\varphi' \to \infty$, the small black hole branch disappears altogether. In this limit, the large black hole branch starts at the origin in the $(W, \tau)$ plane.

\begin{figure}
\begin{center}
 \includegraphics[width=0.9\textwidth]{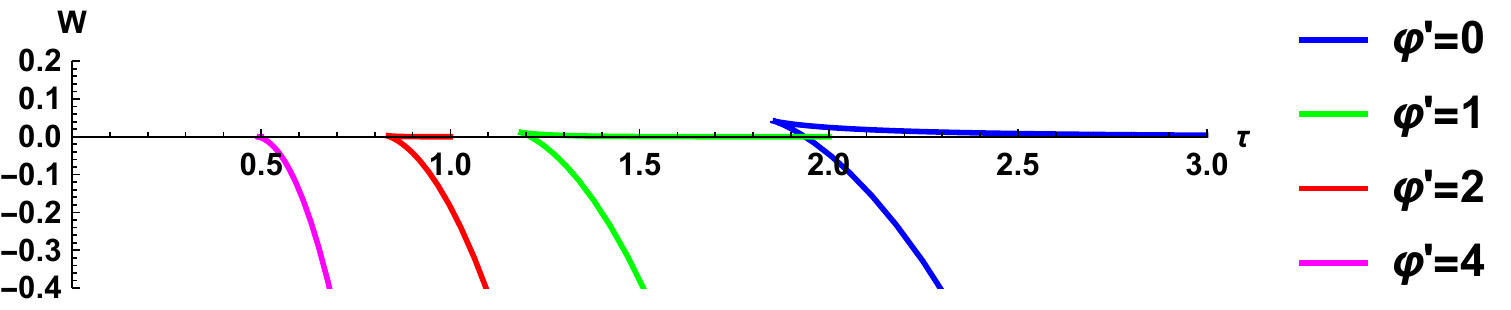}
   \caption{ The BPS free energy $W$ versus the BPS temperature $\tau$ for various values of positive $\varphi'$.  The small black hole branches (the upper branches) have $ W = 0$ at their maximal temperature $\tau_\mathrm{max}$ (\ref{t_max}). Upon increasing $\varphi'$ starting from zero, the $(W_\mathrm{cusp}, \tau_\mathrm{cusp})$ coordinates of the cusp both decrease. The Hawking-Page temperature $\tau_\mathrm{HP}$ where the large black hole branch has $W=0$ also decreases with increasing $\varphi'$.} 
   \label{plot:2}
   \end{center}
\end{figure}

The phase diagram for representative values of positive $\varphi'$ is plotted in figure~\ref{plot:2}. As we  increase $\varphi'$ starting from zero, the $(W_\mathrm{cusp}, \tau_\mathrm{cusp})$ coordinates of the cusp both decrease. The Hawking-Page temperature $\tau_\mathrm{HP}$, where the large black hole branch has $W=0$, also decreases with increasing $\varphi'$. These features can be verified analytically using calculations performed in the previous subsection.

\section{Conclusions}

\label{sec:conclusions}

In this paper, we have developed aspects of AdS$_6$ black hole thermodynamics with focus on the near-BPS limit. Our work generalises the work of Larsen et.~al.~\cite{Larsen:2019oll, Larsen:2020lhg, Ezroura:2021vrt} to six-dimensional setting.  Charged rotating AdS$_6$  and AdS$_7$ black holes with multiple rotation parameters are somewhat different from the gravity perspective \cite{Chow:2007ts, Chow:2008ip}. Typically these solutions are most concisely described in Jacobi-Carter coordinates \cite{Chen:2006xh}, which are less familiar than the more standard Boyer-Lindquist coordinates. This is perhaps one of the reasons that these black holes remain much less studied compared to their  AdS$_5$ cousins \cite{Chong:2005hr}.

In our work, we highlighted that the BPS limit is a two parameter reduction. The two distinct deformations orthogonal to the BPS surface are: (i) increasing the temperature   while keeping the charges fixed, (ii) changing the charges while maintaining extremality such that the BPS constraint  is no longer satisfied. For both these deformations, we proposed a  near-BPS extremization principle that describes the near-BPS regime. The excess entropy together with changes in all potentials are perfectly accounted for via the extremization principle.

The agreements we established show that, at the very least, the near-BPS extremization principle proposed in section \ref{sec:near_BPS_extremization_principle} provides a well-structured packaging of the near-BPS black hole data. Admittedly, the main prescription \eqref{larsen_main_prescription} is somewhat heuristic. In the future, it is desirable to  understand the 
physical configurations ``defined'' by the near-BPS extremization principle directly. In this regard, perhaps relating our analysis to that of section 3.3 of \cite{Cabo-Bizet:2018ehj} will be of help.\footnote{We thank Davide Cassani for emphasising this point to us.}

Microscopic description of the Bekenstein-Hawking entropy formula for BPS AdS$_6$ black holes has been considered in \cite{Choi:2019miv, Crichigno:2020ouj}. We hope that in a near future an  understanding of the near-BPS extremization principle would emerge from the microscopic side. 
 Our results suggest that some sort of non-renormalisation theorem is at work for near-BPS black holes that allows to obtain the entropy and other details from microscopic theories.

Our work offers several opportunities for future research.

We did not compute the on-shell action by evaluating the renormalised on-shell action.  Such a computation can be used to provide a further justification  for our extremization principle. 

The Killing spinors for the  supersymmetric AdS$_6$ black holes have also not been constructed in the literature. It will be nice to fill in this gap too. 

 The phase diagram of near-BPS AdS$_6$ black holes needs to be investigated to understand the condition $\varphi \le 0$.

 In section \ref{sec:phase_diagram}, we saw that there are regular small BPS black holes in AdS$_6$. It is not immediately clear how these black holes are related to the BPS limit of the six-dimensional Cvetic-Youm black holes \cite{Cvetic:1996dt}, which are singular. It will be useful to understand this limit better.

Perhaps the most interesting generalisation of the above considerations would be to the Wu black holes \cite{Wu:2011gq}. These are general non-extremal charged rotating  AdS black holes of five-dimensional  U(1)$^3$ gauged supergravity with three independent charges. From various perspectives these are the most natural black holes to study, but the solutions are so complicated that they remain poorly explored. The study of the type presented above should be manageable and would illuminate further aspects of the much discussed five-dimensional AdS black holes \cite{Aharony:2021zkr, Ntokos:2021duk, Boruch:2022tno}.

 We hope to report on some of these problems in our future work.

\subsection*{Acknowledgments} 
We thank Bidisha Chakrabarty for collaboration in the early stages of this project; and David Chow and  Bidisha Chakrabarty for discussions on the seven-dimensional version of this project. 
We thank Seok Kim, K.~Narayan,  Dharmesh Jain, and especially Davide Cassani for reading through an earlier version of the draft and for suggestions on various points. 
We also thank  Finn Larsen, James Lucietti,  Shruti Paranjape, and Minwoo Suh for email correspondence.  
The work of AV  was supported in part by the Max Planck Partnergroup ``Quantum Black Holes'' between CMI Chennai and AEI Potsdam and by a grant to CMI from the Infosys Foundation.

\bibliographystyle{jhep}

\bibliography{AdS6_final_arxiv_v2}

\end{document}